%% file: ms_final.tex
\documentclass[twocolumn]{emulateapj}

\usepackage{multirow}
\usepackage{makecell}

\shorttitle{Disk--Jet Coupling After a Tidal Disruption Flare}
\shortauthors{Pasham \& van Velzen}

\begin{document}

\title{Discovery of a time lag between the soft x-ray and radio emission of the tidal disruption flare ASASSN-14li: Evidence for linear disk--jet coupling}

%X
 
\author{Dheeraj R. Pasham\altaffilmark{1}}

\affil{Massachusetts Institute of Technology, Cambridge, MA, 02139}
\altaffiltext{1}{NASA Einstein Fellow; dheeraj@space.mit.edu}
\and 
\author{ Sjoert van Velzen\altaffilmark{2}}
\affil{Department of Physics \& Astronomy, Johns Hopkins University, Baltimore, MD 21218}
\affil{Center of Cosmology and Particle Physics, New York University, New York, NY 10003}
\altaffiltext{2}{NASA Hubble Fellow; sjoert@nyu.edu}

\begin{abstract}
The tidal disruption of a star by a supermassive black hole can result in transient radio emission. The electrons producing these synchrotron radio flares could either be accelerated inside a relativistic jet or externally by shocks resulting from an outflow interacting with the circumnuclear medium. Until now, evidence for the internal emission mechanism has been lacking; nearly all tidal disruption flare studies have adopted the external shock model to explain the observed properties of radio flares. Here we report a result that presents a challenge to external emission models: we discovered a cross-correlation between the soft X-ray (0.3-1 keV) and 16~GHz radio flux of the tidal disruption flare ASASSN-14li. Variability features in the X-ray light curve appear again in the radio light curve, but after a time lag of 12$^{+6}_{-5}$ days. This demonstrates that soft X-ray emitting accretion disk regulates the radio emission. This coupling appears to be inconsistent with all previous external emission models for this source but is naturally explained if the radio emission originates from a freely expanding jet. We show that emission internal to an adiabatically expanding jet can also reproduce the observed evolution of the radio spectral energy distribution. Furthermore, both the correlation between X-ray and radio luminosity as well as our radio spectral modeling imply an approximately linear coupling between the accretion rate and jet power.
\end{abstract}

\section{{Introduction}}
The tidal disruption of a star by a massive black hole ($\ga$ 10$^{4}$ $M_{\odot}$) can lead to a spectacular flare that is observable across the entire electromagnetic spectrum. Thermal, i.e., blackbody emission, is detected at optical/UV \citep{Gezari09,vanVelzen10,Arcavi14} and soft X-ray \citep{Bade96,Esquej2008} frequencies and is thought to originate from the tidal debris of the star \citep[e.g.,][; P+17 hereafter]{Pasham17}. Dust reprocessing of this thermal flare can yield a transient signal at mid-infrared wavelengths \citep[e.g.,][]{vanVelzen16b}. 

A handful of candidate stellar tidal disruption flares (TDFs; \cite{Bloom2011}) that have been discovered by their thermal emission are also detected at radio frequencies: ASASSN-14li \citep{vanvelzen2016,alexander2016},  XMMSL1 J0740–85 \citep{Saxton17,Alexander17}, and  IGR J12580+0134 \citep{Irwin15,Perlman17}---although the last is perhaps more likely explained by the activity of a pre-existing active galactic nucleus (see \citealt{Auchettl17} for further discussion). The radio luminosities of these sources are $\sim 10^{38}\,{\rm erg}\,{\rm s}^{-1}$ and fade with a characteristic time scale of a few months. The equipartition energy of the observed radio emission from ASASSN-14li is $\sim 10^{48}$~erg \citep{alexander2016}. This rather low energy and short lifetime leaves open the possibility that such low-luminosity radio flares are common for thermal TDFs. Most radio follow-up observations \citep[e.g.,][]{Bower13,vanVelzen12b} were simply not sensitive enough to detect such events; however see \citet{Blagorodnova17} for a recent counterexample.  

The radio luminosity of the thermal TDFs is over three order of magnitude lower than the radio luminosity of the TDFs that have been discovered by their non-thermal $\gamma$-ray emission \citep{Bloom2011,zauderer2011,Levan11,zauderer2011,burrows2011,Cenko11,Brown15,Pasham15}. The non-thermal X-ray emission of these TDFs is best explained by Doppler-boosted emission caused by a relativistic jet observed at a small inclination. Synchrotron emission due to electrons accelerated in the forward shock of this jet can explain the observed radio light curves \citep{Giannios11,Metzger12,berger2012,Mimica15,Generozov17}. This external emission mechanism is akin to models for radio afterglows of $\gamma$-ray bursts \citep[e.g.,][]{Piran04,Nakar11} or supernovae \citep{chevalier1998}. The isotropic energy of the jet that powers the non-thermal TDFs is $\sim 10^{53}$~erg \citep[e.g.,][]{Mimica15}. 

An external emission model could also explain the observed radio light curves of thermal TDFs. If the accretion onto the black hole exceeds its Eddington limit, we might anticipate the launch of a photon-driven wind \citep[e.g.,][]{olekdiskwind}. When this wind interacts with the circumnuclear medium, shocks can accelerate electrons to yield synchrotron emission  \citep{alexander2016,Alexander17}. An alternative scenario has been proposed by \citet{krolik2016}, who suggests that shocks driven by the unbound stellar debris streams are responsible for producing the synchrotron-emitting electrons. Finally, a third possibility is that the radio emission originates from a forward shock of a relativistic jet that is being decelerated by the circumnuclear medium \citep{vanvelzen2016}. 

We are thus presented with a dichotomy of radio power from TDFs \citep[][]{Generozov17}: low-power thermal TDFs and high-power non-thermal TDFs. However, since the nature of radio emission of the radio-weak TDFs is still under debate, interpreting this dichotomy is difficult. If the radio-weak TDFs are also due to a jet, a unified picture emerges in which all tidal disruptions lead to jet launching \citep[][]{vanVelzen11}, and the observed radio power dichotomy translates into a jet power dichotomy. On the other hand, if radio-weak TDFs are explained by a disk-wind or unbound debris, the observed radio power dichotomy could be explained by the initial conditions that are required for the launch of a relativistic jet \citep[e.g.,][]{Tchekhovskoy13}. 

The goal of this work is to carry out a cross-correlation analysis between the X-ray and the radio light curves of ASASSN-14li to understand the nature of the radio emission of this thermal TDF. Below we first briefly introduce this ``Rosetta stone" TDF. We then present the details of its radio and X-ray observations in Sec.~\ref{sec:data}, followed by the cross-correlation analysis in Sec.~\ref{sec:ccana}. We then discuss our new jet model for the synchrotron emission of this source in Sec.~\ref{sec:syncmodel} and close with a discussion in Sec.~\ref{sec:disc}.

\subsection{ASASSN-14li}
The optical transient \mbox{ASASSN-14li} was discovered by the All-Sky Automated Survey for SuperNovae (ASASSN; \citealt{shappee2014}) on 22 November 2014 or MJD 56983.6 \citep{.pdf}. Within a few weeks it became clear that this source displayed nearly all the known properties of previous optical TDFs: an origin consistent with the nucleus of the host galaxy, an optical spectrum with broad hydrogen and helium emission lines, and a constant blue optical/UV color, corresponding to a temperature of roughly 3$\times$10$^{4}$~K \citep{Holoien16}. 

Two properties are unique to \mbox{ASASSN-14li}: the detection of a luminous, thermal X-ray flare \citep{miller2015} and the detection of a low-luminosity ($\approx$10$^{38}$ erg\,s$^{-1}$ at 16~GHz) radio flare \citep{vanvelzen2016,alexander2016}. The size of the X-ray emitting region as inferred from energy spectral modeling and fast time variability is only a few gravitational radii \citep{miller2015}. This small radius, combined with the fact that the X-ray spectrum is thermal, suggests that the X-rays originate from the innermost regions of the accretion flow \citep{krolik2016}. Because relativistic ejections---which can produce synchrotron radio emission---are also launched from close to the black hole, \mbox{ASASSN-14li} provides a unique laboratory to understand the connection, if any, between the accretion and the ejection of matter near a black hole.

Thanks to its very low redshift, observations with European VLBI Network (EVN) could spatially resolve the radio emission from \mbox{ASASSN-14li} \citep{14lievn2016}. The 5~GHz EVN observations, taken about 200 d after its discovery, revealed a stationary feature with a size of a few milliarcseconds (mas) and a second component $\approx$ 2 parsecs away with a factor of 6 lower in flux. 

The host galaxy of ASASSN-14li (redshift $=0.0206$ or 90 Mpc) is a post-starburst galaxy; TDFs happen preferentially in these rare types of galaxies \citep{Arcavi14,French16}. Integral field spectroscopic observations of the host revealed ionized filaments, similar to ionization nebulae around fading AGNs \citep{Prieto16}. Indeed the detection of low-luminosity radio emission prior to the tidal disruption is best explained by a low-luminosity AGN \citep{Holoien16,vanvelzen2016,alexander2016}.

%++++++++++++++++++++++++++++++++++++++++++++++++++++++++++++++++++++++++++++++++++++++++++++++++++++

\begin{deluxetable}{cccc}
\tablewidth{0pt}
\tablecolumns{4}
\tablecaption{Summary of X-ray Spectral Modeling.}
\tablehead{MJD Range & $N_{\rm H}$ & $T$   & $\chi^{2}$/dof \\ 
& ($10^{22}$ cm$^{-2}$) & (keV)}

\startdata
56991-57014 & 0.106$_{-0.033}^{+0.034}$ & 0.053$_{-0.004}^{+0.004}$ & 1.843/32 \\
57016-57037 & 0.082$_{-0.028}^{+0.032}$ & 0.057$_{-0.004}^{+0.004}$ & 1.224/31 \\
57039-57072 & 0.109$_{-0.031}^{+0.034}$ & 0.054$_{-0.004}^{+0.004}$ & 1.371/32 \\
57075-57137 & 0.094$_{-0.053}^{+0.040}$ & 0.048$_{-0.006}^{+0.007}$ & 0.717/28 \\
57139-57192 & 0.015$_{-0.015}^{+0.028}$ & 0.055$_{-0.005}^{+0.004}$ & 1.416/28 \\
57195-57367 & 0.069$_{-0.037}^{+0.049}$ & 0.044$_{-0.005}^{+0.005}$ & 1.319/24 \\
57370-57834 & 0.176$_{-0.109}^{+0.218}$ & 0.026$_{-0.008}^{+0.008}$ & 1.722/17 \\
\enddata
\tablecomments{$N_{\rm H}$, and $T$ are the best-fit hydrogen column and the blackbody temperature, respectively, of ASASSN-14li's 0.3-1.0 keV average X-ray spectra. The reduced $\chi^2$ along with the degrees of freedom (dof) for the X-ray spectral fit are shown in the last column.}
\label{tab:xana}
\end{deluxetable}

%++++++++++++++++++++++++++++++++++++++++++++++++++++++++++++++++++++++++++++++++++++++++++++++++

\begin{figure*}[t!]
\begin{center}
\includegraphics[height=1.90in, trim=68mm 15mm 30mm 50mm, clip]{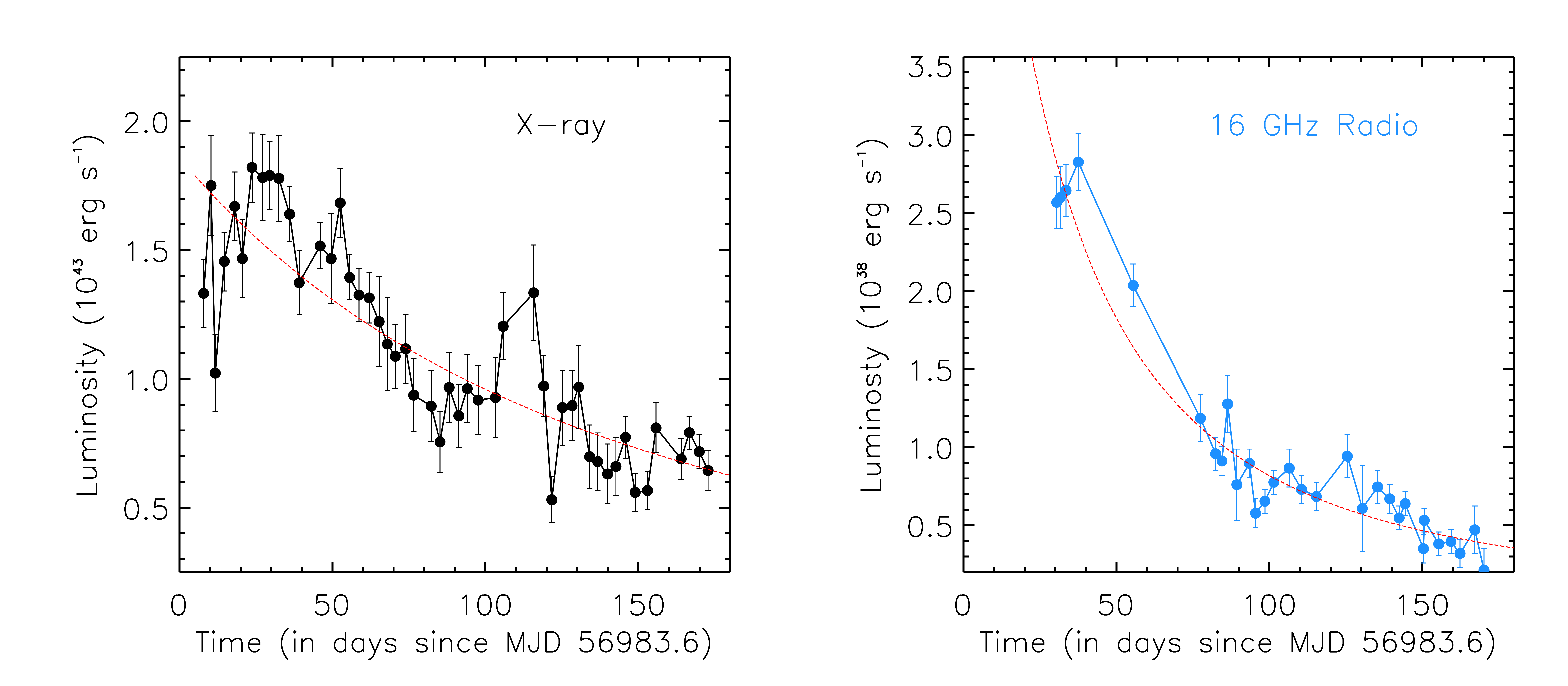}
\includegraphics[height=1.91in, trim=10mm 47mm 30mm 120mm, clip]{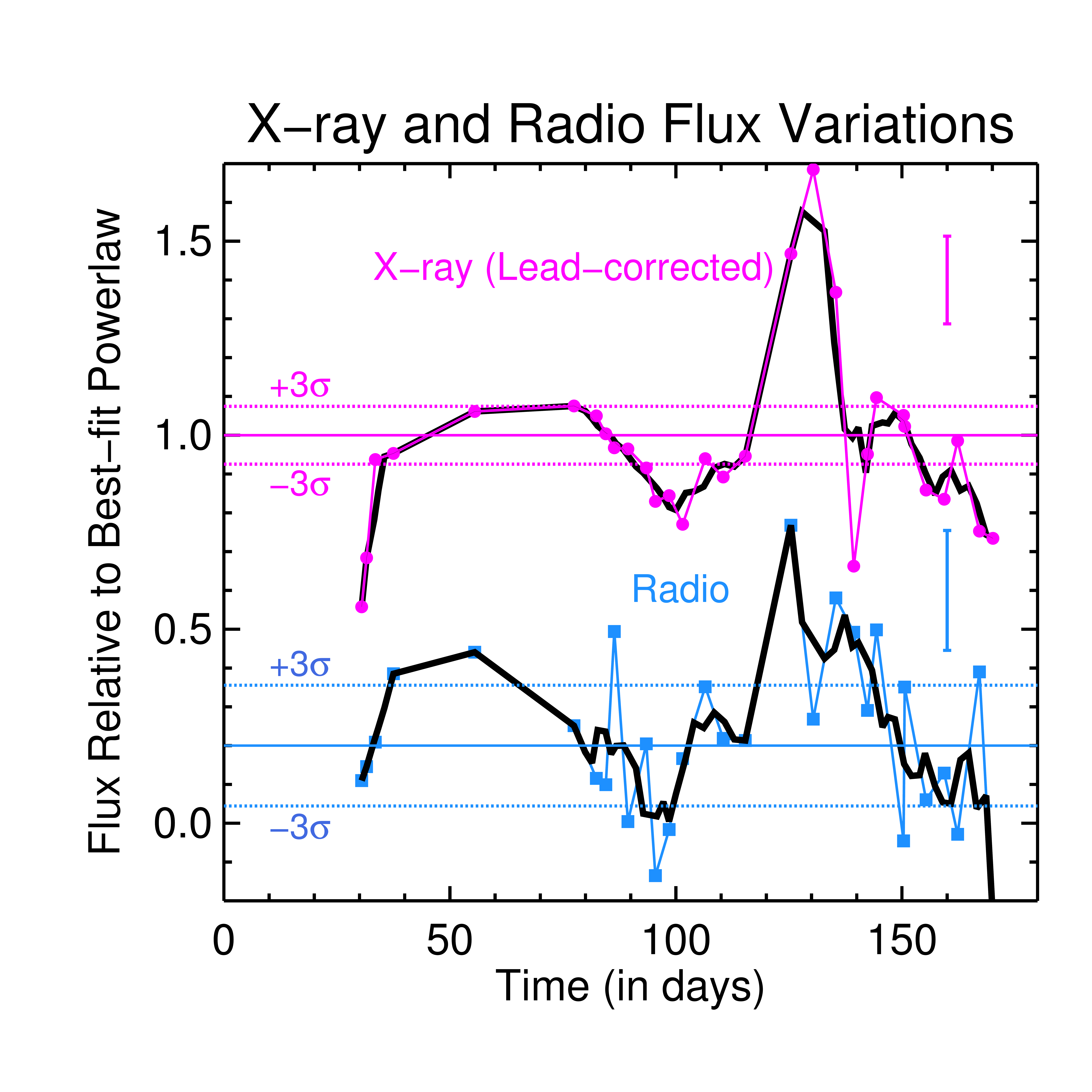}
\caption{De-trending the light curves. {\it Left:} ASASSN-14li's X-ray (0.3-1.0 keV) light curve (magenta data points) along with the best-fit power-law decay model (dashed black curve). {\it Center:} The 15.7 GHz transient radio light curve (blue data points) and its best-fit power-law decay model (dashed black curve). The error bars represent the 1$\sigma$ confidence interval (including the systematic uncertainty of converting from count rate to luminosity). A constant value of 0.24 mJy was subtracted from the observed 15.7 GHz light curve to exclude the non-transient component (see Sec.\ref{sec:radiodata}). {\it Right:} The relative X-ray (magenta) and radio (blue) light curves obtained by dividing their corresponding best-fit power-law model. X-ray and the radio fractional variability amplitudes on top of the power-law decay are 10$\pm$1\% and 16$\pm$1\%,respectively. The mean flux levels are shown by the solid horizontal lines. The radio data has been vertically offset by -0.8. The solid black curves are running averages over a 10 d window (except when the gap between observations in longer than 10 d). Typical 1$\sigma$ uncertainties derived from the left and middle panels are shown as vertical bars. The dashed horizontal lines are the $\pm$3$\sigma$ variability contours derived using the methodology described in \cite{Romero_variability}. The X-ray data points shown here have been interpolated to match with the radio epochs (see Sec.~\ref{sec:visccsub}). Uncertainty on the time lag after fully accounting for these error bars can be found in the middle and right the panels of Fig. \ref{fig:allccs}. }
\label{fig:detrend}
\end{center}
\end{figure*}

%++++++++++++++++++++++++++++++++++++++++++++++++++++++++++++++++++++++++++++++++++++++++++++++++

\section{Radio and X-ray Observations}\label{sec:data}
\subsection{Radio Data}\label{sec:radiodata}
The radio data used in this work were acquired by three different telescopes. The Arcminute Microkelvin Imager (AMI) and the Westerbork Synthesis Radio Telescope (WSRT) provided 15.7 GHz and 1.4~GHz radio data, respectively \citep{vanvelzen2016}. Further multi-epoch radio spectral energy distribution (SED) data were acquired by the very large array (VLA) \citep{alexander2016}. AMI started monitoring \mbox{ASASSN-14li} at 15.7~GHz 31 d after its discovery \citep{vanvelzen2016}. The radio campaign lasted about 140 d with an observing cadence of about one visit every four days. All the radio data used here are as published by van Velzen et al. (2016) and Alexander et al. (2016).

The pre-flare radio flux of \mbox{ASASSN-14li}, based on archival data, is orders of magnitude higher than what is expected from star formation alone. This indicates that a weak AGN was present prior to the flare \citep{Holoien16,vanvelzen2016,alexander2016}. From the plateau in the late-time AMI light curve we infer that the baseline (i.e., non-transient) flux at 15.5~GHz is 0.24~mJy  \citep{Bright17}. To model the observed non-transient flux ($S_{\nu,\rm baseline}$) at other frequencies ($\nu$), one has to account for the source spectrum as well as the difference in angular resolution (or beam) between telescopes and observed frequencies. Here we adopt a simple power-law model with a typical spectral index of $-0.8$ \citep[e.g.,][]{Ker12}:
\begin{equation}\label{eq:baseline}
S_{\nu,{\rm baseline}} = 0.24~\left(\frac{\nu}{15.7~{\rm GHz}}\right)^{-0.8}~{\rm mJy}
\end{equation}
This spectrum is expected if the non-transient radio flux is due to a lobe that was created by the jet that was active prior to the disruption. Our model for the baseline flux (Eq.~\ref{eq:baseline}) is consistent with  the 5 GHz VLA flux that is resolved-out in the EVN observations  \citep{14lievn2016}, suggesting that the region that contains the majority of the baseline emission is smaller than the resolution of the VLA observations at this frequency ($\approx 1"$). Hence the beam difference between AMI and the VLA is expected to have only a modest influence on the observed baseline flux.

\subsection{{X-ray Data Reduction}}\label{sec:xraydata} 

Roughly a week after its discovery, {\it Swift} started monitoring ASASSN-14li with a mean observing cadence of one visit (1-3 ks) every three days.  This cadence was maintained for about 270 days but the monitoring campaign suffered from longer data gaps thereafter because of sun angle constraints.  To ensure that the cross-correlation function (CCF) is not heavily biased by the X-ray data points, we only used the first 180 days of X-ray data (similar in temporal baseline as AMI data) for evaluating the CCF between the X-ray and the 15.7 GHz AMI data. X-ray data were first analyzed by Holoien et al. (2016) and Miller et al. (2015).  However, to properly account for pile-up and estimate the flux we re-analyzed the entire {\it Swift} data, as discussed in detail below.

%++++++++++++++++++++++++++++++++++++++++++++++++++++++++++++++++++++++++++++++++++++++++++++++++
%------------------- FIGURE ----------------------------------------------- FIGURE --------------
%++++++++++++++++++++++++++++++++++++++++++++++++++++++++++++++++++++++++++++++++++++++++++++++++
\begin{figure*}
\begin{center}
\includegraphics[width=6.65in, height=6.0in, angle=0]{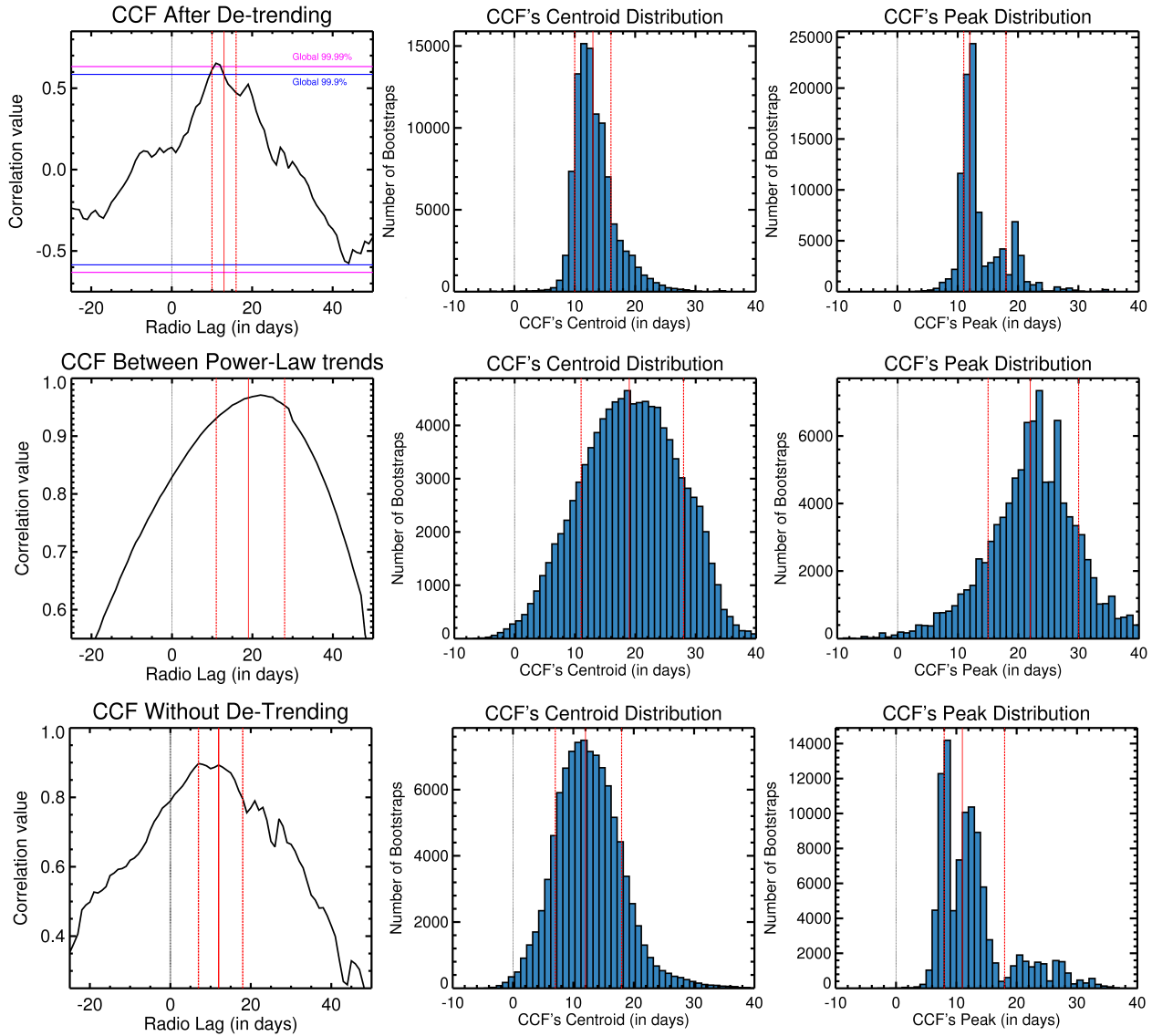}
\end{center}
%\vspace{-.35cm} 
 \caption{ X-ray and radio cross-correlation functions (left) and corresponding centroid (middle) and peak (right) distributions. The top row panels correspond to CCF analysis after de-trending the X-ray and radio light curves. The middle row corresponds to the CCF plots between the best-fit power-law trends of the X-ray and radio light curves. The bottom row shows the CCF plots without de-trending the light curves. {\it The time lag between the X-ray and the radio is evident in all the three cases.} The horizontal blue and magenta lines in the top-left panel are the global 99.9\% and the 99.99\% white noise statistical confidence contours for a search between -25 and 50 days (see Sec. \ref{sec:whitesig}). The centroid and the peak distributions of each CCF are shown in the middle and right panels, respectively. These show the uncertainty in the centroid and the peak of the X-ray--radio lag after taking into account both the measurement and the sampling uncertainties of the light curves. The red dashed vertical lines in these two panels indicate the 1-$\sigma$ deviation away from the median values (solid red lines). The centroids of the CCFs from top to bottom are 13$^{+3}_{-3}$, 19$^{+9}_{-8}$, and 12$^{+6}_{-5}$ days, respectively.}
\label{fig:allccs}
\end{figure*}

%++++++++++++++++++++++++++++++++++++++++++++++++++++++++++++++++++++++++++++++++++++++++++++++++
%------------------- FIGURE 
We reduced the XRT data and extracted pile-up corrected X-ray spectra from each XRT observations following the analysis procedure outlined in P+17. The unner exclusion radius to mitigate pileup was estimated separately from each XRT exposure by modeling the point spread function (PSF) using the method outlined in the {\it Swift}/XRT user guide\footnote{http://www.swift.ac.uk/analysis/xrt/pileup.php}. The inner exclusion radii estimated in this manner are given in the last but one column of Table~\ref{tab:xlong}.

The X-ray spectrum of \mbox{ASASSN-14li} is very soft with almost all counts between 0.3 
and 1.0 keV, and is well-fit with a blackbody model \citep{miller2015}. Therefore, we 
modeled its X-ray spectra with a blackbody function. The individual {\it Swift}/XRT 
monitoring observations lack the necessary signal-to-noise to constrain the blackbody 
temperature and radius. To be able to constrain the model parameters in each XRT observation, 
we carried out an analysis similar to the approach followed by \citet{burrows2011}. 
We grouped data from neighboring epochs until the total counts exceeded 3000 and 
summed spectra using the {\tt ftool} {\tt sumpha}. The corresponding response files, i.e., 
the RMFs (Response Matrix Files) and the ARFs (Ancillary Response Files) were weighed as per 
\mbox{ASASSN-14li}'s counts in the individual epochs and combined using the tasks {\tt addrmf} 
and {\tt addarf}, respectively. This procedure translated to combining anywhere between a few to a few tens of 
neighboring observations. We then fit these averaged X-ray spectra (0.3-1.0 keV) separately 
with a blackbody model defined as {\tt phabs*zashift(phabs*bbodyrad)} in XSPEC 
\citep{arnaud1996}. The best-fit model parameters are listed in  
Table~\ref{tab:xana}. It is evident that the temperature changes were modest during the first 
180 days. This is consistent with Fig.~3 of \citet{miller2015} and more recently with Fig.~5 of \citet{brownjs2017}. 

To obtain the blackbody parameters in each epoch, we modeled the individual XRT spectra with the same blackbody model, but fixed the disk temperature and the 
absorbing column to lie within the error bars of the values of the nearest (in time) averaged spectrum. However, not all individual observations had enough counts to extract an X-ray spectrum. In these cases we assumed that their X-ray spectra have the same shape as that of the observation closest in time to them. Fluxes in these low-count ($<$100) epochs were then estimated by scaling the flux of the nearest observation by the ratio of the PSF-corrected count rates.  The 0.3-1.0~keV fluxes estimated from this approach are listed in Table~\ref{tab:xlong}.

For each XRT pointing we also extracted a mean source count rate corrected for the background, 
the exposure and vignetting using the {\tt ftool} {\tt xrtlccorr}. PSF correction was also performed 
by setting the keyword {\it psfflag = yes}. Exposure maps were also extracted to account for 
bad pixels. The resulting source count rates are shown in Table~\ref{tab:xlong}.

\section{Timing/Cross-correlation analysis}\label{sec:ccana} 
We computed the interpolated cross-correlation function (ICCF; see \citealt{gaskell1987,white1994,peterson2004} and the references therein) between the 15.7 GHz radio and the X-ray light curves. As explained in Sec.~\ref{sec:radiodata}), we subtracted the non-transient emission at 15.7 GHz, but we stress that the observed correlation and the lag (below) are independent of the value of the baseline flux.

Welsh (1999) argued that long-term trends in the light curves can sometimes bias the ICCF 
analysis and thus de-trending\footnote{ De-trending refers to removing the long-term trend from the light curves.} with a smooth function can improve the ability to recover the 
correlation and the lag between the two time series under consideration. Therefore, we first 
de-trended both the X-ray and the radio light curves with a power-law decay model. The two 
light curves along with their best-fit decay model of the form A$\times$(t-t$_{0}$)$^{-\alpha}$ 
(where A, t$_{0}$, and $\alpha$ are the model parameters, and t is the time in days) are shown 
in  Fig.~\ref{fig:detrend} (left and middle panels).  The best-fit index value, $\alpha$, and the $\chi^2$/degrees of freedom for the X-ray light curve are 2.5$\pm$2.4 and 127/46, respectively. For the radio the corresponding values are 1.9$\pm$0.1 and 62/27, respectively. It should be noted that the purpose of modeling here is to only make an estimate of the long-term trend. Moreover, the index value and the time at peak (t$_{0}$) are degenerate (but see Sec.~\ref{sec:linearcoup} where we estimate t$_{0}$ by modeling the evolution of radio spectra). The ICCF was then computed between the residual light curves (data minus the best-fit model). We also experimented with other de-trending models, viz., a linear 
model, and a second order polynomial model. The ICCFs in all these cases 
showed clear evidence for a radio lag around 13 days. Because of the theoretical expectation that 
the bolometric TDF light curves should decay as a power-law \citep{rees1988,Lodato09}, which is consistent 
with most X-ray observations \citep{komossa2002}, we adopted the analysis corresponding to the 
power-law decay model. In contrast, P+17 used a bending-powerlaw model, but we stress that the cross-correlation between the X-ray and radio light curves is independent of the de-trending model (See Fig. \ref{fig:allccs}). We computed the ICCFs using the steps outlined in P+17 (left panels of Fig.~\ref{fig:allccs}).

\subsection{{ Visually assessing the cross-correlation}}\label{sec:visccsub}
To allow a visual assessment of the strength of the cross-correlation, we placed the X-ray and radio light curve on top of each other (Fig.~\ref{fig:detrend}, right panel). We first displaced the observed X-ray light curve by 13 d, the median of the ICCF's centroid (see Sec.~\ref{sec:cccd}). We then interpolated the X-ray light curve onto the time values of the radio data. The interpolated X-ray light curve (magenta) is overlaid on the 15.7 GHz radio data (blue) in the right panel of Fig.~\ref{fig:detrend}. It is clear that they both exhibit the same variability features. The black curves are a running mean of the data points in a 10 d window.  Adapting the procedure outlined by \cite{Romero_variability} we also extracted the 3$\sigma$ variability contours. These are shown as dashed horizontal lines in the right panel of Fig.~\ref{fig:detrend}.

\subsection{{ Statistical Significance Contours}}\label{sec:whitesig}
%\subsubsection{White Noise Statistical Significance}\label{sec:whitesig}
The X-ray--radio ICCF in Fig.~\ref{fig:allccs} clearly shows a peak centered around 13 d\footnote{The notation 
throughout the paper is such that a positive lag in the ICCF implies that the radio lags behind 
the X-rays.}. To assess the statistical significance of this peak, we estimated the global 99.9\% and 99.99\% Gaussian white noise confidence contours as follows. First, we extracted M values from a Gaussian 
distribution with a mean of zero and a standard deviation equal to the standard deviation of the 
observed de-trended X-ray light curve. Here, M is the number of data points in the observed X-ray 
light curve. By assigning these M values to be the flux values at the times of the observed X-ray 
light curve, we constructed a synthetic de-trended X-ray light curve sampled exactly as the 
observed X-ray light curve. This synthetic X-ray light curve was then cross-correlated with the 
de-trended radio light curve exactly as the observed ICCF. This procedure was repeated 10$^{6}$ 
times to construct 10$^{6}$ synthetic CCFs. We then constructed a distribution of all CCF values between -25 and 50 d. From this distribution we extracted the global 99.9 and 99.99\% significance levels (see the left panel of Fig.~\ref{fig:allccs}).

\subsection{Uncertainty on the Lag's Centroid and Peak}\label{sec:cccd}
We extracted the errorbars on the peak and the centroid lag values using the Flux Randomization (FR) and the Random Subset Selection (RSS) methods as described in Peterson (2004). These account for the measurement and the sampling uncertainties of both the light curves. We estimated the distribution of the centroid (top-middle panel of Fig.~\ref{fig:allccs}) and the peak (top-right panel) of the ICCF. The centroid was estimated using all correlation values greater than 0.8 times the peak. 

Finally, we also repeated the entire cross-correlation analysis with the X-ray count rate light 
curve (instead of flux) and the resulting ICCF is consistent with the ICCF of the X-ray flux versus 
the radio flux. This is not surprising as the X-ray spectral changes were only modest (see 
Table~\ref{tab:xana}) and therefore count rate serves as a good indicator of the flux.

%++++++++++++++++++++++++++++++++++++++++++++++++++++++++++++++++++++++++++++++++++++++++++++++++
%------------------- FIGURE ----------------------------------------------- FIGURE --------------
%++++++++++++++++++++++++++++++++++++++++++++++++++++++++++++++++++++++++++++++++++++++++++++++++
\begin{figure}
\begin{center}
\includegraphics[width=0.45 \textwidth, trim=65mm 2mm 20mm 0mm,clip]{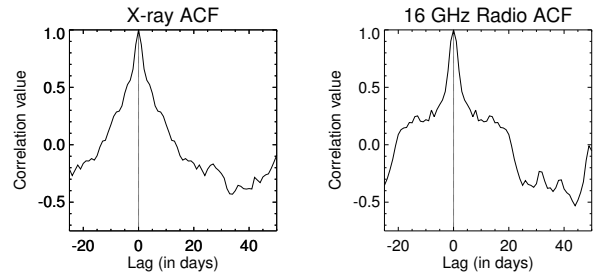}
\caption{X-ray and radio auto-correlation functions (ACFs). The ACF of the X-ray and the radio light curves are shown in the left and the right panels, respectively. Both are obtained using the same CCF parameters as in Fig.~\ref{fig:allccs}. The difference in the two ACF shapes is only an artifact of the difference in the sampling of their respective light curves (see Sec.~\ref{sec:acf}).}
\label{fig:ACFs}
\end{center}
\end{figure}
%++++++++++++++++++++++++++++++++++++++++++++++++++++++++++++++++++++++++++++++++++++++++++++++++
%------------------- FIGURE ----------------------------------------------- FIGURE --------------
%++++++++++++++++++++++++++++++++++++++++++++++++++++++++++++++++++++++++++++++++++++++++++++++++

\subsection{{Time Lag between the X-ray and Radio Trends}}
We also extracted an ICCF between the best-fit power-law trends of the X-ray and the 15.7~GHz radio light curves. These trends are indicated as red dashed curves in the left and middle panels of Fig.~\ref{fig:detrend}. Because the highest cross-correlation power in these trends is at the lowest frequency, one might expect the CCF of these two declining light curves to simply peak at zero lag. Instead, we find that the strongest cross-correlation value is at a finite lag (see Fig.~\ref{fig:allccs}). The best-fit lag centroid value is 19$^{+9}_{-8}$ d and is consistent with the lag obtained from only the short-time scale fluctuations of the light curves, i.e.,  the CCF of the de-trended light curves (top panels of Fig.~\ref{fig:allccs}). This strongly suggests that not only are the variations in the radio and the X-rays are correlated but the entire radio-emitting region is regulated by the X-ray engine. 

%++++++++++++++++++++++++++++++++++++++++++++++++++++++++++++++++++++++++++++++++++++++++++++++++++

\subsection{CCF without de-trending}
Finally, to establish that de-trending is not causing the correlation and the time lag, we also extracted an ICCF and its corresponding centroid and peak distributions between the observed 15.7 GHz radio and soft X-ray fluxes. These are shown in the bottom panels of Fig.~\ref{fig:allccs}. As expected the correlation and the lag are evident. In addition to ruling out any de-trending related artifacts this establishes that the entire radio and soft X-ray flux of ASASSN-14li are correlated with each other.

\subsection{{ Auto-correlation Functions}}\label{sec:acf}
A cross-correlation function is a convolution of the auto-correlation function (ACF) with the 
transfer function. A lag is real only if it originates from the transfer function and not from 
the ACF itself. To ensure that the lag seen in Fig.~\ref{fig:allccs} did not originate from either of the 
X-ray or the radio ACF we evaluated both and show them in Fig.~\ref{fig:ACFs}. 
Clearly, both the ACFs are centered on zero lag and thus assert that the observed lag originates 
from the transfer function. 

The difference in the shapes of the two ACFs is due to the different sampling of the two light 
curves. Whether the X-ray light curve is lead-corrected and interpolated onto the radio epochs or 
vice versa, the resulting ACFs are similar (which is expected for two light curves that are highly correlated).

\begin{figure}
\centering
\includegraphics[width=0.44 \textwidth, trim=6mm 10mm 0mm 12mm, clip]{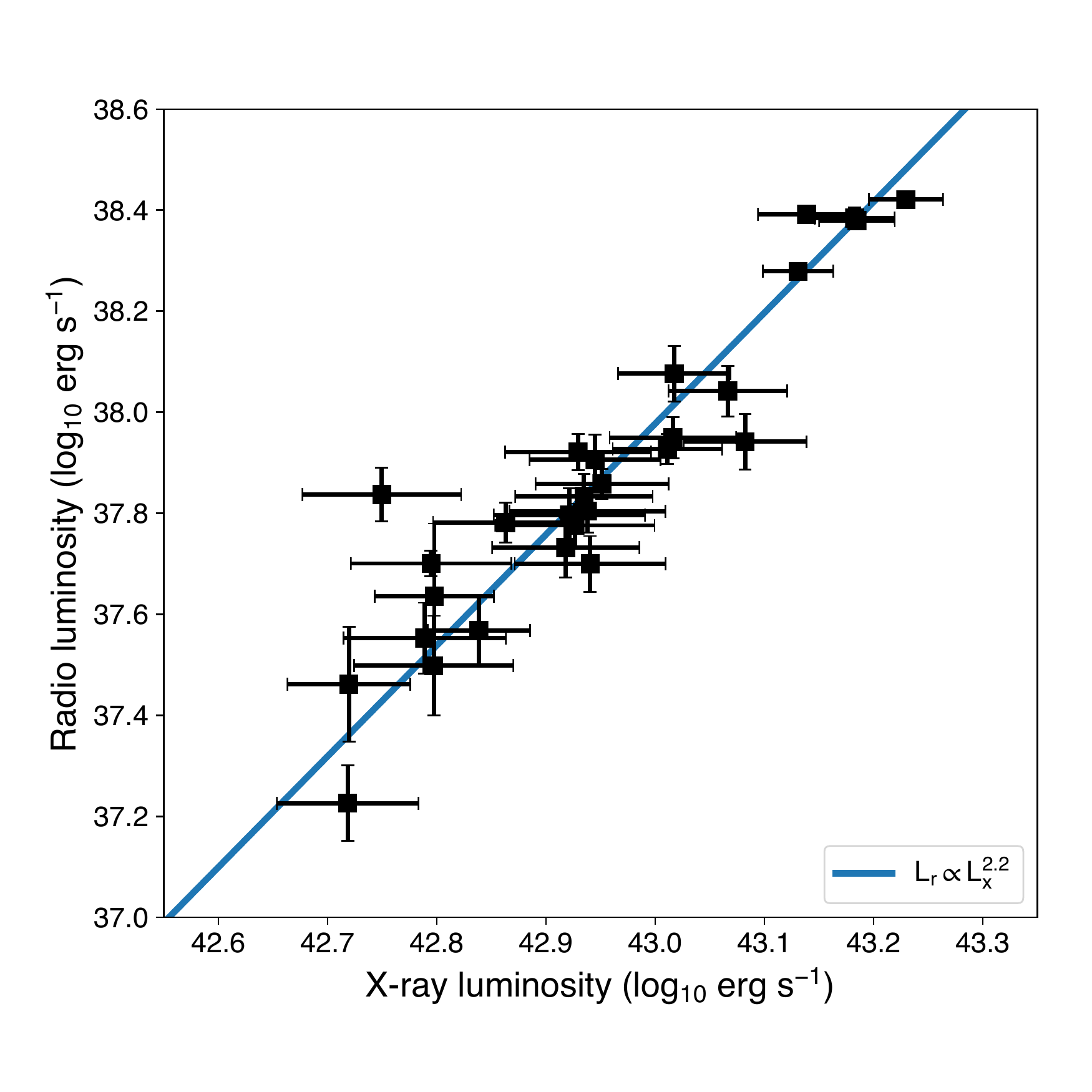}
\caption{ {X-ray and radio luminosity.} Here we show the lead-corrected X-ray luminosity and the 16 GHz radio luminosity (combining both the AMI and the VLA observations, both with the non-transient flux subtracted). The correlation between the luminosity in these two wavelengths can be described as $L_{\rm radio} \propto L_{\text{ X-ray}}^{2.2}$. The reduced $\chi^2$ of the best-fit power-law relation is 0.6. This index suggests that the accretion and jet power of ASASSN-14li are linearly coupled (see Sec.~\ref{sec:linearcoup}).}
\label{fig:LrLx}
\end{figure}

%++++++++++++++++++++++++++++++++++++++++++++++++++++++++++++++++++++++++++++++++++++++++++++++++
%------------------- FIGURE ----------------------------------------------- FIGURE --------------
%++++++++++++++++++++++++++++++++++++++++++++++++++++++++++++++++++++++++++++++++++++++++++++++++

\subsection{{Correlation between X-ray and Radio Luminosity}}
After correcting for the lag between the X-ray and the radio light curves, we compared the contemporaneous luminosities at these two frequencies. Since the X-ray data are sampled with a higher cadence, we interpolated the X-ray light curve onto the time stamps of the radio data. Following the ``Nuker method'' \citep{tremaine2002}, we fit a power-law to the X-ray versus radio luminosity by minimizing the following merit function: 
\begin{equation}\label{eq:merit}
    \chi^2(t_{\rm lag}) = \sum \frac{(L_{\rm radio}(t)+ a - b L_{\rm X\textnormal{-}ray}(t+t_{\rm lag}))^2  }
    {\sigma_{\rm radio}^2 + b^2 \sigma_{\rm X\textnormal{-}ray}^2}
\end{equation}
Here $b$ is the X-ray--radio power-law index and $a$ is the normalization; $L_{\rm X\textnormal{-ray}}(t)$ and $L_{\rm radio}(t)$ denote the logarithms of the X-ray and the radio luminosity, respectively, as a function of time. $\sigma_{\rm X\textnormal{-}ray}$ and $\sigma_{\rm radio}$ are the logarithms of the X-ray and radio measurement uncertainties, respectively. The X-ray light curve is interpolated to a time $t+t_{\rm lag}$, with $t_{\rm lag}$ being the observed CCF lag. We can estimate the statistical uncertainty on the best-fit power-law index under the assumption that Eq.~\ref{eq:merit} follows $\chi^2$ statistics \citep{tremaine2002}. To estimate the systematic uncertainty on the power-law index, we sampled the observed distribution of lags (Fig.~\ref{fig:allccs}, bottom middle panel) and obtained a distribution of best-fit values of $b$.

We find the following best-fit power-law index $b=2.2\pm0.2 {(\rm statistical}) \pm 0.3 ({\rm systematic})$, with a reduced $\chi^2$ of 0.6. The data and this power-law relation are shown in Fig.~\ref{fig:LrLx}. If no lag is applied to the X-ray light curve, the scatter in the luminosity--luminosity relation is significantly larger and the reduced $\chi^2$ of the best-fit power-law is 2.0. An interpretation of the X-ray--radio luminosity relation is presented in Sec.~\ref{sec:linearcoup}.

\section{Spectral/Synchrotron analysis}\label{sec:syncmodel}
\subsection{Evolution of the  Peak Frequency}\label{sec:fpeak}
Using the formulae of synchrotron emission and self-absorption (e.g., \citealt{Pacholczyk70}) and an electron energy ($E_{e}$) distribution that follows a power law ($N_e dE_e\propto E_{e}^{\rm -p} dE_e$), \citet{marscher1985} derived the dependence of the peak radio flux ($S_{\rm peak}$) on the frequency at peak ($\nu_{\rm peak}$) for an adiabatically expanding region in a conical jet to be,
\begin{equation}\label{eq:Marscher85}
    S_{\rm peak} \propto \nu_{\rm peak}^{\frac{\rm 10(p-1)}{\rm 7p+8}}
\end{equation}
The above equation (Eq. 20, in \citealt{marscher1985}) assumes that the magnetic field strength ($B$) scales with the radius, $r$, (perpendicular to the jet axis) as $B \propto r^{-1}$ which is the case for a conical jet \citep{readhead1978,FB1995}. Using an electron power-law index (p) between 2 and 3, the above equation then translates to
\begin{equation}\label{eq:mg}
    S_{\rm peak} \propto \nu_{\rm peak}^{0.57 \pm 0.12} \quad.
\end{equation}
On the other hand, the peak synchrotron flux of a cloud that is expanding radially is given by Eq. 17b of \citet{vanderlaan1966}:
\begin{equation}
    S_{\rm peak} \propto \nu_{\rm peak}^{\frac{\rm 7p+3}{\rm 4p+6}} \quad,
\end{equation}
or, for $2<p<3$,
\begin{equation}\label{eq:vdl}
     S_{\rm peak} \propto \nu_{\rm peak}^{1.27 \pm 0.06} \quad.
\end{equation}
The difference in the two cases is due to the scaling of the magnetic field and the particle energy with radius. For an adiabatically expanding blob in a conical jet geometry, the energy of a relativistic electron ($E$) falls off with radius as $E \propto r^{\rm -2/3}$ (see \citealt{Marscher1980}), while for a  cloud of electrons that is adiabatically expanding in a spherical geometry, we have $E \propto r^{-1}$ (see \citealt{vanderlaan1966})\footnote{The scaling of the particle energy with source radius can be derived from the ideal gas equation for relativistic particles as follows. The ideal gas equation of state for relativistic particles is $Pr$ $\propto$ $n_{e}^{\rm 4/3}$, where $Pr$ and $n_{e}$ are the pressure and the electron density in the gas, respectively. We have $Pr = n_{e} \left<E\right>$, where $\left<E\right>$ is the mean energy of an electron in the gas, hence $\left<E\right> \propto n_{e}^{\rm 1/3}$. For a spherical cloud of relativistic electron gas, the electron density falls off with radius as $n_{e} \propto r^{\rm -3}$. Thus $\left<E\right> \propto r^{\rm -1}$ for a spherical geometry. However, for an expanding jet, $n_{e} \propto r^{\rm -2}$ and thus $\left<E\right> \propto r^{\rm -2/3}$.}.

Eqs.~\ref{eq:mg} and \ref{eq:vdl} can be compared to the observed evolution of ASASSN-14li's radio SED to discriminate between an expanding jet and an expanding spherical cloud. \citet{alexander2016} modeled each radio SED with a single volume that is in equipartition. Using their values of the peak flux and frequency we find $ S_{\rm peak} \propto \nu_{\rm peak}^{0.61 \pm 0.04}$.

\begin{figure}
\begin{center}
\includegraphics[width=0.48\textwidth, trim=10mm 0mm 10mm 0mm, clip]{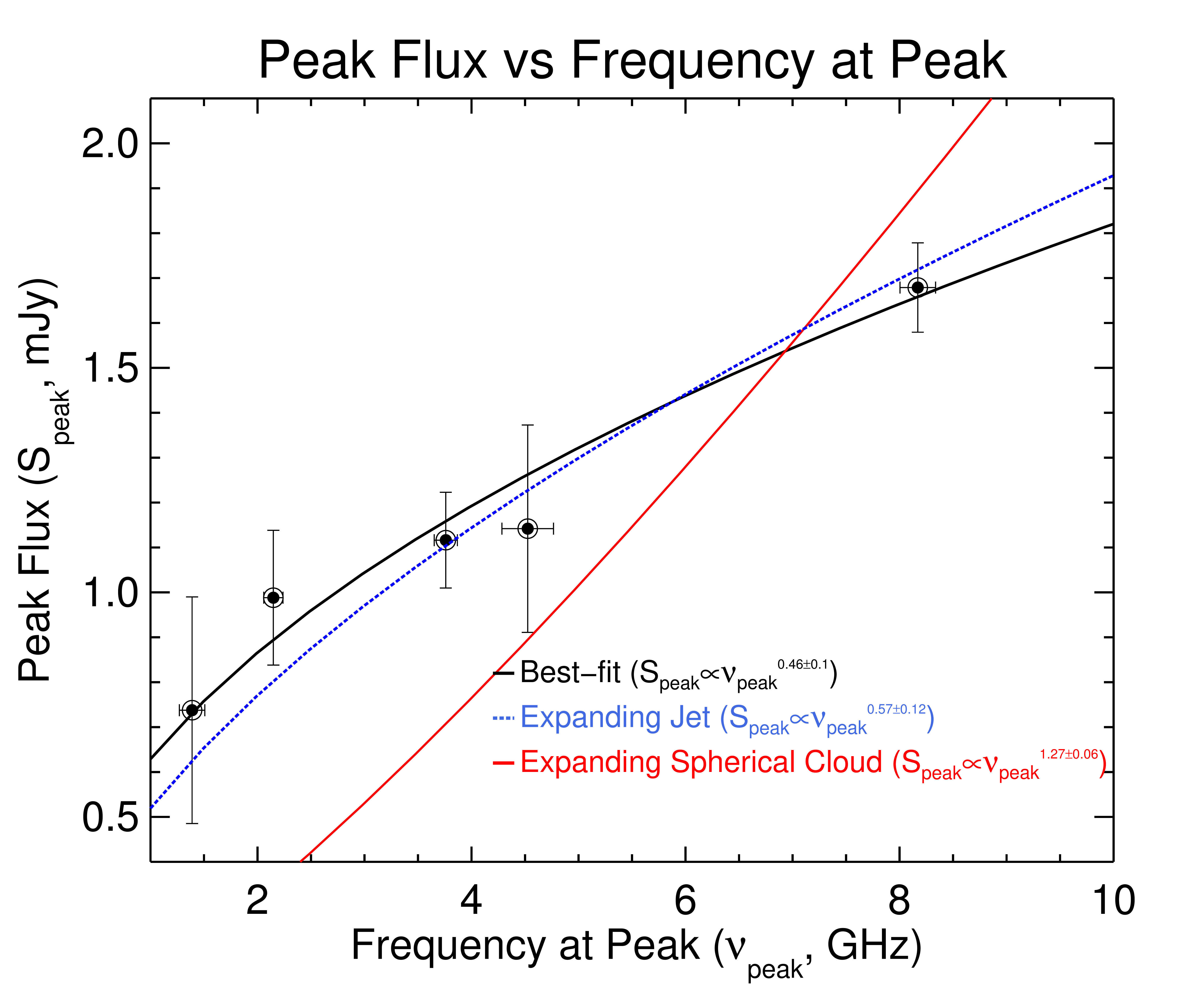}
\caption{Radio SED evolution: conical jet compared to a spherical cloud. Evolution of the peak radio flux vs. the frequency at peak (black data points). The best-fit power-law index of 0.46$\pm$0.1 is shown by a solid black curve while the index for an adiabatically expanding jet \citep{Marscher1980} and an adiabatically expanding spherical cloud \citep{vanderlaan1966} are shown by the dashed blue and the red curves, respectively.}
\label{fig:SEDevo}
\end{center}
\end{figure}

In order to extract the peak flux and the frequency at peak in a model-independent way, we fit the radio SEDs with a simple bending power-law function of the form:
\begin{equation}
     S_{\nu} = \frac{S_{0}\nu^{-\alpha_{1}}}{1+\Big(\frac{\nu}{\nu_{\rm bend}}\Big)^{(\alpha_{2}-\alpha_{1})}}
\end{equation}
Here, $S_{0}$ and $\nu_{\rm bend}$ are the normalization and the frequency at which the SED turns over, respectively. $\alpha_{1}$ and $\alpha_{2}$ are the spectral index for frequencies ($\nu$) much smaller than and much larger than $\nu_{\rm bend}$, respectively. We fixed the low-frequency slope ($\alpha_{1}$) to be equal to the value for synchrotron self-absorption ($\alpha_1=-2.5$). Using the five of the best sampled SEDs from \citet{alexander2016}, i.e., those taken on 2015 Jan 6/13, 2015 Mar 13, 2015 April 21/22, 2015 June 16/21 and 2015 Aug 28/Sept 8/11, we find that the observed peak flux ($S_{\rm peak,obs}$) and the frequency at peak ($\nu_{\rm peak,obs}$) evolve as
\begin{equation}
     S_{\rm peak,{\rm obs}} \propto \nu_{\rm peak,{\rm obs}}^{0.46 \pm 0.1}
\end{equation}
This observed scaling disfavors a uniformly expanding single spherical cloud model and is consistent with the bulk of the radio emission originating from a conical geometry that is adiabatically expanding, see Fig.~\ref{fig:SEDevo}. 

%++++++++++++++++++++++++++++++++++++++++++++++++++++++++++++++++++++++++++++++++++++++++++++++++
%------------------- FIGURE ----------------------------------------------- FIGURE --------------
%++++++++++++++++++++++++++++++++++++++++++++++++++++++++++++++++++++++++++++++++++++++++++++++++

%++++++++++++++++++++++++++++++++++++++++++++++++++++++++++++++++++++++++++++++++++++++++++++++++
%------------------- FIGURE ----------------------------------------------- FIGURE --------------
%++++++++++++++++++++++++++++++++++++++++++++++++++++++++++++++++++++++++++++++++++++++++++++++++

\subsection{Single-zone Models}\label{sec:singlezone}
Consider a spherical region of radius $R$ that is uniformly filled with magnetic fields ($B$) and relativistic electrons  with an energy distribution $N_{e} d\gamma_e = K \gamma_{e}^{\rm -p} d\gamma_e$, where $\gamma_{e}$ and $K$ are the electron Lorentz factor and the normalization, respectively. When observed at a distance $D$, the observed synchrotron flux of this region is given by
\begin{equation}\label{eq:sync}
S_{\nu} = \delta^{2} \frac{R^{2}}{4D^{2}} \frac{\epsilon_{\nu'}}{\kappa_{\nu'}} (1-e^{-\kappa_{\nu'} R}) \quad.
\end{equation}
Here $\delta$ is the Doppler factor of the region, and $\nu$ is the observed frequency. The synchrotron absorption ($\kappa_{\nu'}$) and emission ($\epsilon_{\nu'}$) coefficients depend on the magnetic field, the electron energy distribution, and the frequency in the rest-frame of the jet, $\nu'=\nu/\delta$. We solve for the normalization of these coefficients by assuming equipartition between the energy in the electrons and the magnetic field, $B^{2}/8\pi = \int N_{e} d\gamma_{e} $. Equipartition implies a minimum in the total energy of the system (i.e., the sum of the energy in the magnetic  field and the synchrotron-emitting particles). Observations of the lobes of radio galaxies \citep{croston2005,Kataoka2005,hardwood2016} as well as AGN jets \citep{burbidge1956,readhead1978,herrnstein1997,falcke1999} provide strong evidence that equipartition is commonly reached for these sources. We confirmed that for $\gamma_{\rm max}\gg \gamma_{\rm min}$, our implementation of the synchrotron formalism (Eq.~\ref{eq:sync}) gives the same results as the fitting formula for synchrotron emission given in \citet{chevalier1998}.

As expected based on earlier work \citep{alexander2016,krolik2016}, we find that the synchrotron emission from a single region (Eq.~\ref{eq:sync}) provides a reasonable fit to the radio SEDs of ASASSN-14li. For this fit we adopted $\gamma_{\rm min}=1$, p$=2.2$ and $\gamma_{\rm max}=10^{4}$, but we stress that the resulting magnetic field and radius are only weakly dependent on these assumptions. For $\delta=1$ (i.e., a non-relativistic outflow), this single-zone synchrotron model yields $B\approx 0.1$~G and $r\approx 4\times 10^{16}$~cm for the observations of April 2015. For this magnetic field, the  synchrotron cooling time at 10~GHz is $\sim 10$~yr, which supports the assumption that $ \gamma_{\rm max}\gg \gamma_{\rm min}$. For this single-zone model, we find that the equipartition magnetic field scales with source size as $B\propto R^{-1.2}$.

The hotspots or the forward shock of a jet or outflow can be modeled using a single-zone equipartition model. However, establishing the observed cross-correlation in a single-zone model proves to be very difficult. First of all, the light crossing time of this region grows from $\approx 10$~d to $\approx 30$~d during the period of the 15.7~GHz monitoring observations, thus exceeding the duration of the cross-correlation lag. A second problem of establishing an X-ray--radio correlation within a single region is the long synchrotron cooling time at 16~GHz ($\sim 10$~yr). Over the course of the 16~GHz monitoring observations ($\approx$180 d), the region cannot radiate the energy it has received and thus the relative amplitude of fluctuations in the radio light curve due to fluctuations in the X-ray light curve should decrease with time. 

To summarize, while a single-zone equipatition model for ASASSN-14li can reproduce the observed radio SEDs, its size is likely to be too large to produce variability on a 10~d timescale. Moreover, due to the long synchrotron cooling time, the addition of new energy to  has a negligible effect on the radio flux.  This leads us to consider a freely expanding jet as the source of the observed radio emission. 

%++++++++++++++++++++++++++++++++++++++++++++++++++++++++++++++++++++++++++++++++++++++++++++++++
%------------------- FIGURE ----------------------------------------------- FIGURE --------------
%++++++++++++++++++++++++++++++++++++++++++++++++++++++++++++++++++++++++++++++++++++++++++++++++
% trying to keep these figures on one page, with the schematic on the left column. 
\begin{figure}
\begin{center}
\includegraphics[width=3.0in, trim=70mm 5mm 50mm 2mm,clip]{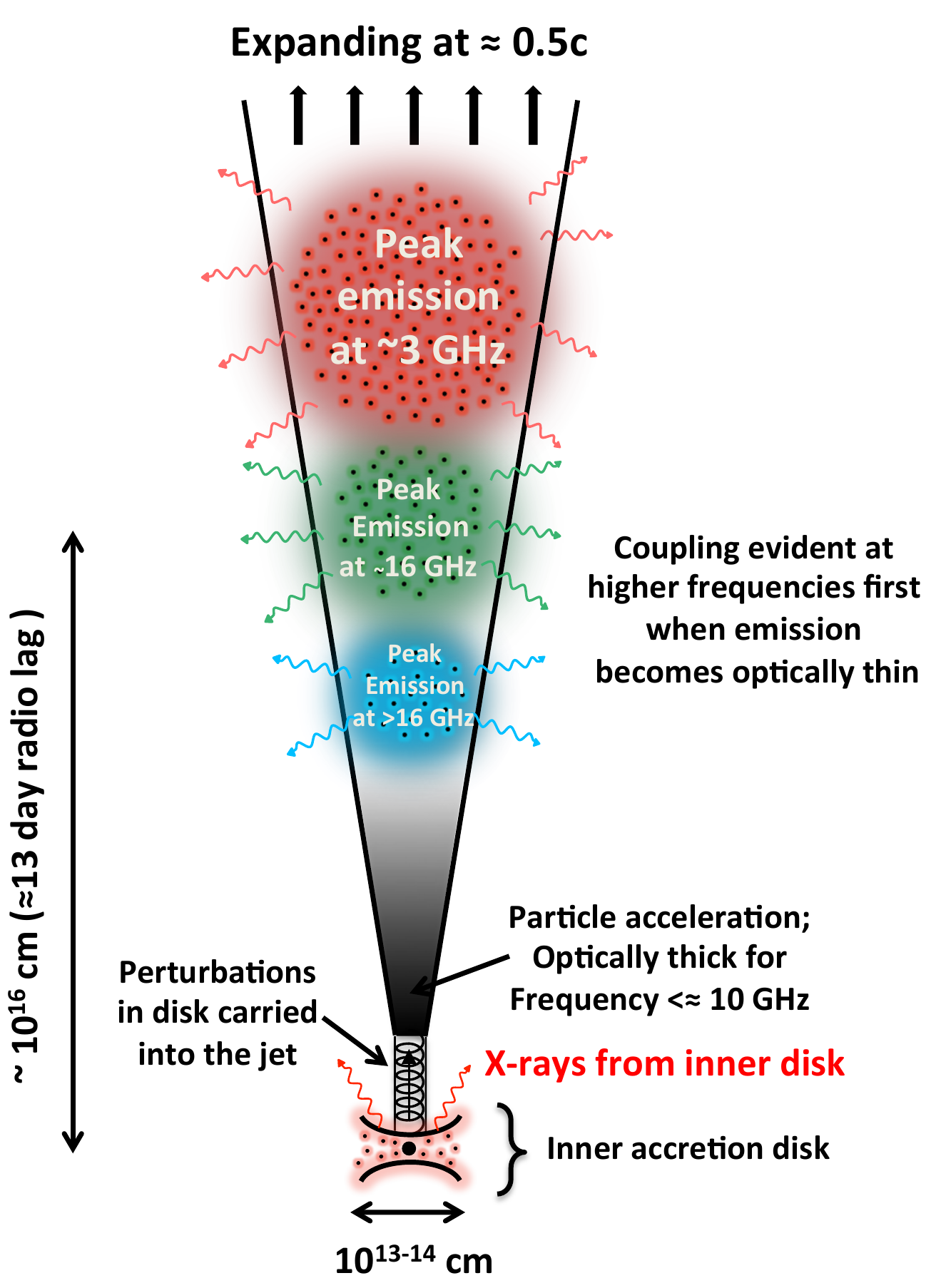}
\end{center}
\caption{Schematic of the proposed jet model for the tidal disruption flare ASASSN-14li. Shown here is a snapshot of the jet at the end of the radio monitoring observations (June, 2015). The X-ray--radio correlation and the radio spectral evolution can both be explained as follows. First, perturbations in the accretion rate are manifested as X-ray flux variations and, via the disk--jet coupling, lead to perturbations in the jet power. The jet power is used to accelerate electrons, which produce synchrotron emission. As the synchrotron radiating electrons are swept further along the jet axis, they start to cool adiabatically. When their emission becomes optically thin to self-absorption at 15.7 GHz, at $\sim 10^{16}$~cm from the black hole (about 13 d later), the observed X-ray--radio correlation emerges. Applying our jet model to the radio observations of ASASSN-14li (Fig. \ref{fig:jetmodel}), we estimate the jet flow velocity at these radii to be about 0.5$c$.}
\label{fig:schematic}
\end{figure}
%++++++++++++++++++++++++++++++++++++++++++++++++++++++++++++++++++++++++++++++++++++++++++++++++
%------------------- FIGURE ----------------------------------------------- FIGURE --------------
%++++++++++++++++++++++++++++++++++++++++++++++++++++++++++++++++++++++++++++++++++++++++++++++++
\begin{figure}
\begin{center}
\includegraphics[width=0.5\textwidth, trim=6mm 10mm 0mm 12mm, clip]{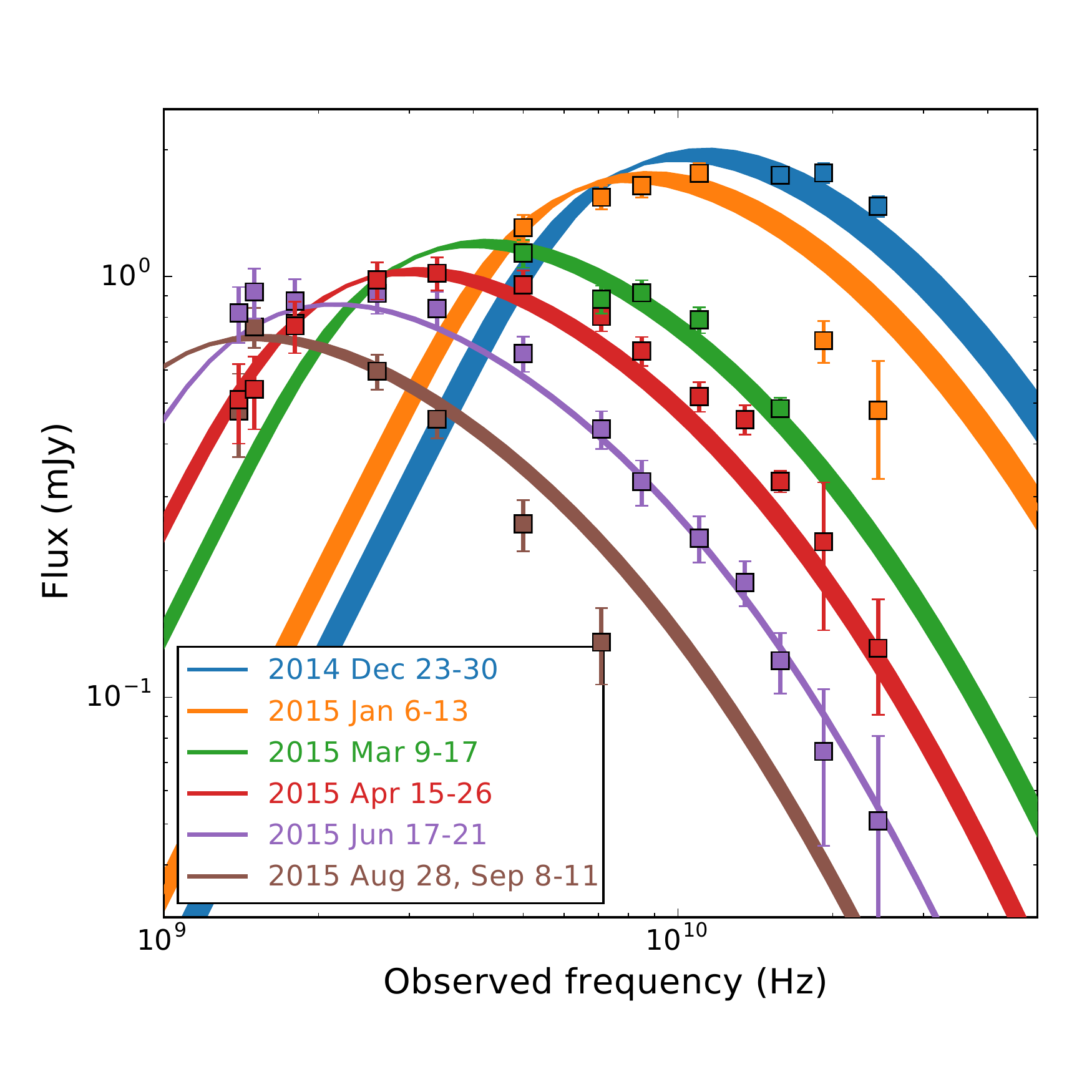}
\caption{Adiabatic jet model and  multi-frequency radio observations of ASASSN-14li. This jet model is a superposition of synchrotron-emitting spheres, each expanding with the same velocity in a conical jet geometry. The electrons in each region in the jet cool adiabatically, which yields the decrease of the peak luminosity with time. Data points with the same colors are semi-simultaneous (epochs are labeled in the legend). The width of each model curve indicates the range of the predicted flux due to the temporal spread of the observations.}
\label{fig:jetmodel}
\end{center}
\end{figure}
%++++++++++++++++++++++++++++++++++++++++++++++++++++++++++++++++++++++++++++++++++++++++++++++++
%------------------- FIGURE ----------------------------------------------- FIGURE --------------
%++++++++++++++++++++++++++++++++++++++++++++++++++++++++++++++++++++++++++++++++++++++++++++++++

\subsection{Adiabatic Jet Model}\label{sec:jetmodel}
A conical equipartition jet model is widely used to explain the properties of compact radio cores of AGN and X-ray binaries \citep{blandfordkonigl1979,FB1995,falckeII1995,crumley2017}. In this model, electron acceleration happens internal to the jet (e.g., via collimation shocks) and equipartition is established at each radius in the jet that lies beyond the nozzle where the jet is accelerated. 

Summing the optically thin synchrotron emission in the model for AGN jet cores yields the characteristic flat spectrum, $S_{\nu}\propto \nu^{0}$. However, the peak of the radio SED of  ASASSN-14li decreases with time, $S_{\rm peak} \propto \nu_{\rm peak}^{0.5} \propto t^{-0.5}$. This suggests that we are observing the adiabatic evolution of electrons that have been heated prior to becoming optically thin to synchrotron self-absorption \citep{marscher1985}. Besides the evolution of the peak frequency (see Sec.~\ref{sec:fpeak}), additional evidence for this adiabatic evolution is the apparent exponential turnover at $\nu \approx 15$~GHz in the radio SEDs (see Fig.~\ref{fig:jetmodel}). This turnover is most clearly seen in the radio data taken on 2015 Aug 28, Sep 8-11 (see Fig.~\ref{fig:jetmodel}). While this can be explained by synchrotron cooling, matching the cooling time at 16~GHz to the dynamical time requires a magnetic field that is two orders of magnitude higher than the observed equipartition value. In other words, the observed high-frequency break in the radio spectra can be explained by synchrotron cooling, but only if the particles were accelerated in a region with a magnetic field that is higher than the equipartition value. This can be established if the acceleration happened downstream in a jet (i.e., closer to the black hole), where the magnetic strength is larger. To include the effect of synchrotron cooling on the spectral shape, we allow the maximum Lorentz factor of the electrons, $\gamma_{\rm max}$, to be a free parameter in our jet model.

\input{table_jetfit_newsub.tex}

To predict the light curve in an adiabatic jet model, we use a superposition of non-overlapping spheres in a conical geometry (Fig.~\ref{fig:schematic}), each with a flux given by their magnetic field and radius (Eq.~\ref{eq:sync}). If each sphere receives the same amount of jet power, the total flux (i.e., the contribution from all the spheres) yields the well-known flat spectrum, $S_{\nu} \propto \nu^{0}$. Following \citet{vanderlaan1966} and \citet{marscher1985} we account for adiabatic cooling of electrons via the normalization of the electron energy distribution
\begin{equation}\label{eq:Kcooling}
K(z>z_{0}) = (z/z_{0})^{ 2(1-p)/3 }.
\end{equation}
Here $z$ is the distance measured along the jet axis and $z_{0}$ is the distance from the black hole where electrons are no longer accelerated and the jet starts to cool. Adding this cooling term to Eq.~\ref{eq:sync}, we retrieve the scaling of \citet{marscher1985} for the peak flux with frequency at peak of an adiabatic jet (Eq.~\ref{eq:Marscher85}). 

Since we have a rapid decrease in the accretion power, we expect that the jet power ($Q_j$) downstream from the jet head ($z_{\rm head}$) will decrease. We model this with a power-law scaling,
\begin{equation}\label{eq:cQ}
Q_{j}(z,t) =  (z/z_{\rm head}(t))^{c_{Q}} \quad,
\end{equation}
with $c_{Q}$ a free parameter. A second free parameter of our jet model is the scaling of the magnetic 
field along the jet axis
\begin{equation}\label{eq:cB}
B(z,t) = B_{0} (z/z_0)^{c_B} \times Q_{j}^{1/2}(z,t) \quad.
\end{equation}
If no magnetic energy is lost and the jet can freely expand to yield a conical geometry we expect 
$c_{B}=-1$ \citep{Blandford74,blandfordkonigl1979,FB1995}. We stress that $c_{B}$ and $c_{Q}$ are not degenerate; $c_{B}$ parameterizes how the radio flux from every regions in the jet changes as it expands while $c_{Q}$ determines how the jet power scales relative to the peak jet power carried by the jet head. 

The next two free parameters of our model concern the jet dynamics. Motivated by the inference that a linear growth provides a good description of the equipartition radius as a function of time in a single-zone synchrotron model \citep{alexander2016,krolik2016}, we assume a linear growth of jet head with time 
\begin{equation}\label{eq:vjet}
z_{\rm head}(t) = v_{\rm jet}\, \delta (t-t_{0}) \quad,
\end{equation} 
with $v_{\rm jet}$ the rest-frame jet velocity and $t$ is measured in the observer frame. 

The last two free parameters in our model are the jet opening angle, $\phi = r/ z$, and the inclination, $i$. When the inclination is known, the Doppler factor can be computed using the jet velocity, $\delta = 1/(\gamma_{\rm jet} (1- \beta_{\rm jet} \cos i)$. We also compute the emission from the counter jet, observed at $i-\pi$, but its contribution is sub-dominant for most inclinations. 

Eqs. \ref{eq:sync}--\ref{eq:vjet} together yield our jet model. This model has nine free parameters,  ($v_{\rm jet}$, $t_{0}$, $B_{0}$, $z_{0}$, $c_{Q}$, $c_{B}$,  $\gamma_{\rm max}$, $\phi$, and $i$),  compared to the 82 observations of the radio flux (the combination of the WSRT, AMI, and VLA data). For comparison, determining the velocity in a single-zone equipartition model requires six  parameters: $v$, $t_0$, $B_0$, $i$, $\gamma_{\rm max}$, plus a factor to account for the geometry of the emitting region.  

If no adiabatic cooling is included, $z_{0}$ is simply a normalization constant with no physical meaning, i.e., it can be set to any value. However, when cooling is included (Eq.~\ref{eq:Kcooling}), $z_{0}$ can be thought of as the distance from the black hole where adiabatic cooling becomes important. We set $z_{0}=10^{15}~{\rm cm}$, but we stress that this choice has no effect on the inferred jet dynamics or magnetic field scaling along the jet axis. When adiabatic cooling starts, the jet internal energy scales as $z^{-8/3}$ \citep{crumley2017}. Hence for our choice of $z_{0}$, the jet has lost a factor $\sim 10$ of its energy by the time of the first radio observations. (In principle, $z_{0}$ can be constrained if we include a self-consistent treatment of  synchrotron cooling in our jet model; while this is beyond the scope of this work, we do note that $z_{0}\ll 10^{15}$~cm can be ruled-out since this yields a $\gamma_{\rm max}$ that is too low to explain the observed radio SEDs). 

We fixed the jet inclination at its a priori most-likely value, $i=60^{\circ}$, but we will explore less probable inclinations below. Finally, the opening angle of the jet is poorly constrained by the radio observations alone. To allow the new jet to be freely expanding, it likely has to remain within the volume swept clear by the jet that was active prior to the tidal disruption. We therefore adopt $\phi=1/10$ as a fiducial value and also consider the best-fit jet parameters for larger/smaller opening angles. 

In our fit for the parameter of the adiabatic jet model, we enforced a 5\% minimum statistical uncertainty on the radio data. This avoids putting too much weight on the early VLA observations, where the statistical uncertainty almost certainly exceeds the variance due to the limitations of our simple model. To subtract the non-transient radio flux, we use Eq.~\ref{eq:baseline}. If we instead adopt the non-transient flux that was used in the analysis of \citet{alexander2016}, we infer similar values for the free parameters of our model.

We use a least-squares fit to estimate the parameters of our jet model; the results are summarized in Table~\ref{tab:jetmodel}. The reduced $\chi^2$ of the best fit is 3.6. To approximately include the variance due to the limitations of our model into the derived parameters, we multiplied the statistical uncertainty of the best-fit parameters by $\sqrt{3.6}$.

We find that the inferred jet velocity depends on the assumed opening angle, for $1/15<\phi<1/5$ we obtain $0.3<v_{\rm jet}/c<0.7$. The inferred values of the other parameters are essentially independent on the opening angle. Since the jet Lorentz factor is modest, the effect on the jet inclination on the best-fit parameters is relatively small (e.g., for $\phi=1/10$ and $i=90^\circ$, $v_{\rm jet}=0.6c$, while the same jet opening angle observed at $i=0$ yields $v_{\rm jet}=0.5c$). We find that 80\% of the total jet flux is reached when the first four synchrotron-emitting spheres are summed, with the first sphere (i.e., the jet head) contributing about 30\% of the total flux.  

The power-law index that sets the scaling of the magnetic field strength along the jet axis is a free parameter in our model. We find $c_{B} = -1.02\pm 0.03$, in excellent agreement with the expected $B\propto z^{-1}$ scaling for a conical jet geometry and conservation of magnetic energy \citep{blandfordkonigl1979,FB1995}. By extrapolating to  $z=0$, we estimate that the jet was launched near the second week of June, 2014. 

When the magnetic field is known, our jet model can be used to estimate the optical depth of each zone in the jet. Since the cross-correlation can only be observed when the synchrotron emission is optically thin, our prediction for the radius where $\tau=1$ at 16~GHz provides a consistency check on our jet model. This radius yields an estimate of the communication velocity between the disk and the region from where the majority of the 16~GHz flux originates. The communication velocity can be estimated as $v_{\rm lag}/c =$ $\gamma^{-1} (\tau_{\rm lag} c/z_{\rm lag} + \cos(i))^{-1}$, with $\tau_{\rm lag}$  the observed delay between the X-ray and the radio light curves. Using our best-fit value for the magnetic field at MJD=57018 (which corresponds to the peak of the first correlated feature in the two light curves, about one month into the 16~GHz monitoring campaign, see Fig. \ref{fig:detrend}), this estimate of $v_{\rm lag}$ is consistent with jet velocity predicted by our model (see Table~\ref{tab:jetmodel}).

\subsection{{Coupling between Accretion Rate and Jet Power}}\label{sec:linearcoup}
Under the assumption of a constant expansion velocity, our estimate of the jet power scaling obtained in the previous subsection can be translated to a scaling of the jet power with time. Along the jet-axis, $Q_j\propto z^{1.2\pm 0.4}$. Under the assumption of a constant expansion velocity, we thus obtain $Q_j\propto t^{-1.2\pm 0.4}$. Interestingly, this relation is consistent with the slope of the observed X-ray flux decay, $L_{\text{X-ray}}\propto (t-t_0)^{-1.7\pm 0.1}$ (here we fixed the time normalization, $t_0$, to our estimate of the time when the jet was launched). This X-ray flux decay index is also close to the expected fallback rate of the stellar debris, $t^{-5/3}$ \citep{phinney1989}. Because the thermal X-ray energy spectrum suggests an efficient accretion disk we expect the X-ray luminosity to be proportional to the mass accretion rate, $L_{\text{X-ray}} \propto \dot{m}$ (e.g., see Fig. 1 of \citealt{Sadowski11}; \citealt{Abramowicz13}; \citealt{Lodato11}). From the evolution of the radio spectral energy distribution we thus find evidence that the jet power decays in concert with the accretion rate.   

A second piece of evidence for linear jet--disk coupling follows from the correlation between the X-ray and radio luminosity. Because in section~\ref{sec:ccana} we concluded that entire radio flux is correlated with the X-rays (see Fig. \ref{fig:allccs}) we can use this  correlation to estimate the coupling strength between the mass accretion rate and the jet power (see also \citealt{Bright17}). 

Using Eq. (7) of \cite{heinz_sunyaev03} it can be seen that the optically thin synchrotron emissivity, $j_{\nu}$, at a given radio frequency ($\nu$) for a power-law distribution of electrons with index p is given as

\begin{equation}
    j_{\nu} = J_{p} K B^{\frac{\rm p+1}{2}}\nu^{-\frac{\rm p-1}{2}}
\end{equation}

\noindent where $J_{p}$ is a constant weakly dependent on p, and $B$ is the magnetic field strength, and $K$ is defined as before (see \ref{sec:singlezone}).

Thus, at a given radio frequency 
\begin{equation}\label{eq:lrad} 
    j_{\nu} \propto L_{\rm radio} \propto KB^{\frac{\rm p+1}{2}}
\end{equation}
where $L_{\rm radio}$ is the radio luminosity at $\nu$. The jet power scales with 
the magnetic field strength as $Q_{j} \propto B^{2}$ and, under the assumption of equipartition
$K \propto B^{2}$. Combining these two relations, we find (\citealt{heinz_sunyaev03}, their Eq.~16) 

\begin{equation}\label{eq:lrad_Qj}
    L_{\rm radio} \propto Q_{j}^{1 + \frac{\rm p+1}{4}}
\end{equation}
For, 2 $<$ p $<$ 3 the index is 1.875$\pm$0.125.

From the observed X-ray and radio luminosities, we have $L_{\rm radio} \propto L_{\text{X-ray}}^{2.2\pm0.3}$. Combining these into Eq. \ref{eq:lrad_Qj} results in

\begin{equation}
    Q_{j} \propto \dot{m}^{1.2\pm0.2}
\end{equation}
We thus see that the accretion and the jet power follow a roughly linear coupling. A caveat is that Eq.~\ref{eq:lrad_Qj} is valid for a single region in the jet. While the 16 GHz emission is dominated by the radius where the jet becomes optically thin to self-absorption, other regions also provide a sub-dominant contribution to the 16~GHz flux. Correcting for this requires a more complete jet model than the toy model used in this work.

\section{Discussion}\label{sec:disc}

Our main conclusions, in order, are as follows.
\begin{enumerate}
    \item We detected a correlation, significant at greater than the 99.99\% level, between the soft X-ray and the 15.7 GHz radio variability of the thermal TDF ASASSN-14li. The radio emission lags the X-rays by about 12 d (Fig.~\ref{fig:allccs}). 
     \item The cross-correlation is inconsistent with external emission models (i.e., shocks driven into the circumnuclear medium) since in such models the radio emission is expected to evolve independently of the accretion rate or fallback rate. 
    \item We propose that the electrons responsible for the observed synchrotron emission are accelerated inside a jet which provides a natural vehicle to couple the radio-emitting region with the X-ray emitting region (Fig.~\ref{fig:schematic}).
    \item Emission from a cloud of electrons that is adiabatically expanding in a conical jet geometry provides a good match to the observed evolution of the radio SEDs  (Figs.~\ref{fig:SEDevo} \& \ref{fig:jetmodel}). 
    \item Our jet model correctly predicts the observed time lag between the radio and X-ray light curves. 
    \item The observed scaling between the X-ray and the 16 GHz radio flux (Fig.~\ref{fig:LrLx}) suggests that the accretion and the jet power are roughly linearly coupled (see Sec. \ref{sec:linearcoup}).
    \item From our jet model we obtain the jet power as a function of radius (Table~\ref{tab:jetmodel}), again finding a scaling that is consistent with a linear disk--jet coupling (see Sec. \ref{sec:linearcoup}). 
\end{enumerate}

Below we discuss a few implications of these conclusions. First, we discuss our results in context of P+17, followed by a comparison of ASASSN-14li to sources on the fundamental plane, and then briefly remark on the implications for jet physics. 

\subsection{Connection to the Optical/UV--X-ray cross-correlation}\label{sec:p17connec}
P+17 discovered that the bulk of the optical/UV emission from ASASSN-14li is produced roughly 32 d ahead of the X-rays. They suggested that energy and perturbations from debris stream self-interactions could produce optical/UV emission and their corresponding fluctuations, respectively. These are then carried down to the inner accretion region where they modulate the X-rays. Combining this with the radio lag reported here suggests that the optical/UV emission does not originate from the jet. In other words, radio does not originate from the same site as the optical/UV light. 

The observation that the fluxes in a wide range of the electromagnetic spectrum are correlated with X-rays---and thus with each other---leads to the following account that ties together the multi-wavelength properties of ASASSN-14li. The UV/optical emission is produced first, at the location where the debris streams intersect. Matter then falls to the center to form a compact, X-ray emitting accretion disk. Perturbations to energy supply at the location of the optical/UV region are carried into the X-ray disk and ultimately show up as variability in the radio-emitting jet.

\begin{figure}
\begin{center}
\includegraphics[width=0.45\textwidth, trim=20mm 8mm 2mm 40mm, clip]{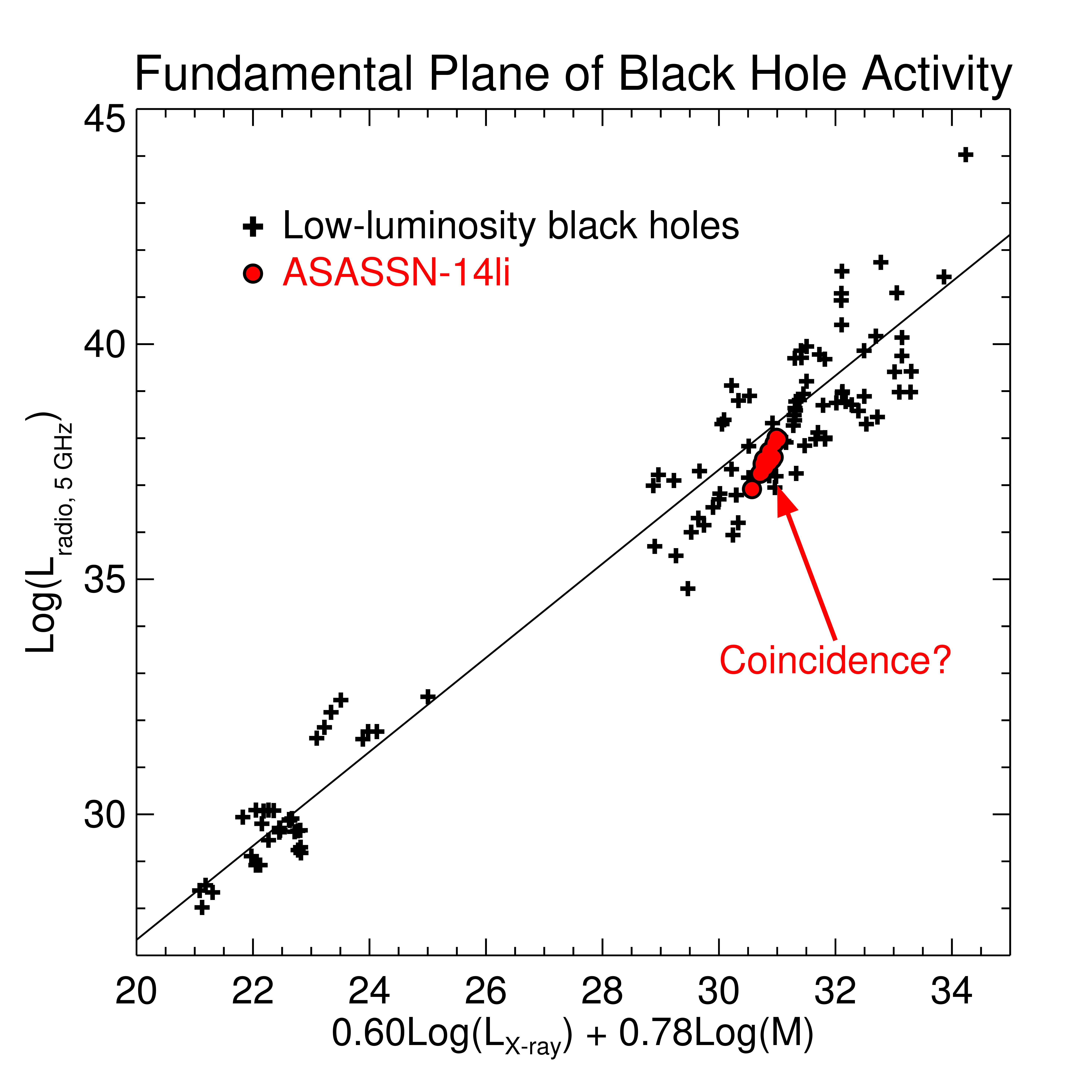}
\caption{{ASASSN-14li on the fundamental plane of black hole activity.} The black data (crosses) show the fundamental plane of black hole activity as derived in \cite{merloni2003} with their best-fit relation. Observations of ASASSN-14li data are shown in red (filled circles). L$_{\text{X-ray}}$, M, and L$_{\rm radio, 5GHz}$ are the 2-10 keV \mbox{X-ray} luminosity, and the black hole mass in units of solar mass, and the 5 GHz radio luminosity, respectively. Because ASASSN-14li is a very soft X-ray source \citep{miller2015}, its X-ray luminosity is estimated in the 0.3-1.0 keV energy range. Moreover, it should be cautioned that the 0.3-1.0 keV luminosity of ASASSN-14li only represents a small fraction of the accretion luminosity and is different from the typical 2-10 keV used for low-luminosity black holes.}
\label{fig:fundamentalplane}
\end{center}
\end{figure}

\subsection{Comparison to AGN and X-ray Binary Jets}
Black holes accreting at only a few percent of their Eddington limit are known to exhibit 
accretion--jet coupling \citep{merloni2003,falcke2004,gallo2003}. For some supermassive black holes it has been possible to detect a time lag on the order of a few tens of days between the radio emission from the compact jet core and the hard X-ray ($>$2 keV) emission \citep{marscher2002,bell2011,chatterjee2011}. These observations all point to a coupling between black hole jets and accretion disks.

Work by \citet*{merloni2003} and \citet*{falcke2004} has shown that for black holes in the ``hard-state" (see \citealt{RemillardMcClintock06} for a definition of this accretion state), the hard X-ray (2-10 keV) luminosity, the 5 GHz radio luminosity and the black hole mass are correlated, thus spanning a plane. This relation---the fundamental plane of black hole activity---extends all the way from stellar to supermassive black holes, ranging over roughly seven orders of magnitude in black hole mass (see  Fig.~\ref{fig:fundamentalplane}). 

To place ASASSN-14li on the fundamental plane of black hole activity, the 5 GHz radio luminosity was estimated at each 15.7 GHz epoch by interpolating the radio SEDs using our best-fit jet model. It is evident that the soft X-ray luminosity and the 5 GHz radio luminosity of ASASSN-14li fall very close to the fundamental plane. Allowing the mass to vary in between $10^{5}$ to $10^{7}$ $M_{\odot}$ the observations of ASASASN-14li do not move significantly away from the plane. 

Our finding that the X-ray and radio properties of ASASSN-14li are consistent with the fundamental plane of black hole activity could be considered surprising. First of all, the slope of the X-ray--radio correlation of ASASSN-14li (2.2, Fig.~\ref{fig:LrLx}) is much steeper than slope of the plane (0.6), at a fixed black hole mass. Furthermore,  the X-ray emission from ASASSN-14li is thermal and soft (0.3-1 keV), thus likely originating from an radiatively efficient accretion flow \citep{Abramowicz13}; the X-ray luminosity of sources on the fundamental plane is dominated by non-thermal emission (in the 2-10 keV band). These hard X-rays presumably originate from an X-ray corona rather than a radiatively efficient accretion disk. It has been suggested that the X-ray corona can either be an inefficient inner accretion flow \citep{yuancui2005} or the base of the jet, among other possibilities \citep{markoff2005}. In either scenario, the X-ray emission site for these  black holes is physically different from ASASSN-14li. The close match to the fundamental plane could simply be a coincidence, although we note that the 2--10 keV output of the powerful jetted TDF {Swift}~J1644+57 is also consistent with the fundamental plane \citep{millersw1644radio2011}.

\subsection{A Jet Power Dichotomy?}
A linear coupling between the jet power and accretion power is well-established for radio-loud quasars \citep{Rawlings91,falckeII1995,vanVelzen15}. The hard X-ray light curve of the powerful jetted TDF {Swift}~J1644+57 closely matches the power-law decay expected for the fallback rate of the stellar debris \citep{Levan16}. Since the X-rays of this source almost certainly originate from the base of a relativistic jet \citep[however see,][]{Kara16}, these observations suggest that  {Swift}~J1644+57 may also displays a linear disk--jet coupling. Combining this with our discovery of a linear coupling between the accretion rate and the jet power for ASASSN-14li, may lead one to conclude that a similar jet engine operates for both radio-loud quasars and jets from TDFs such as ASASSN-14li and {Swift}~J1644+57. However this unified picture is challenged by the difference in radio luminosity between these two TDFs.

The difference in the jet energy of {Swift}~J1644+57 and the equipartition energy of ASASSN-14li is about 4 orders of magnitude \citep{Alexander17}. Relativistic Doppler boosting is not a solution to explain this difference since the late-time radio emission of {Swift}~J1644+57 is most likely isotropic \citep{Mimica15}. We further note that the host galaxies of these events have a similar central black hole mass, $M\sim 10^6\, M_\odot$, as inferred from the host galaxy properties \citep{Levan16,Wevers17}. 
Since our jet model for ASASSN-14li requires adiabatic cooling to explain the evolution of the SED, the true jet energy of this source is likely to be an order of magnitude higher than the equipartition energy inferred from the radio SED (see Sec.~\ref{sec:jetmodel}). However, to be consistent with observed radio SED, the correction due to adiabatic cooling cannot be arbitrarily large. We thus conclude that the difference in isotropic jet power between {Swift}~J1644+57 and ASASSN-14li is 2-3  orders of magnitude. 

The difference in jet power could be explained by a difference in jet efficiency, i.e., the conversion of accretion power to jet power. Parameters that may affect this efficiency are black hole spin and the magnetic field near the horizon. Alternatively, to uphold a unified scenario in which all TDFs have a similar jet efficiency, the amount of accretion energy available for jet production must be higher for {Swift}~J1644+57. This difference could be established by the orbit of the star. Consider a deeply plunging orbit (i.e., a pericenter much smaller than the tidal radius). These events likely yield  rapid circularization, and thus accretion, of the stellar debris \citep{Dai15}. The result is a longer super-Eddington accretion phase compared to stars that are disrupted close to the tidal radius. If the jet is powered only when the accretion rate is super-Eddington \citep[e.g.,][]{Coughlin14}, deeply plunging tidal disruptions could thus yield more powerful jets.

While a freely expanding jet appears to provide the only self-consistent explanation for both the observed radio SEDs and the X-ray--radio correlation of ASASSN-14li, an uncomfortable feature of this model is the requirement of a low external particle density on a scale of $10^{16}$~cm from the black hole. This can be reconciled if the newly launched TDF jet propagates along the same axis as the jet that was operating during the AGN activity prior to the stellar disruption. Thermal TDFs by black holes that had no active jet prior to the disruption may therefore not produce month-long radio flares. This could explain the radio non-detection in a recent nearby TDF \citep{Blagorodnova17} where a radio flare similar to ASASSN-14li would have been detectable.

%++++++++++++++++++++++++++++++++++++++++++++++++++++++++++++++++++++++++++++++++++++++++++++++++

\acknowledgments
{Acknowledgments}: This work is based on observations made with {\it Swift}, a mission that is managed and controlled by NASA's Goddard Space Flight Center (GSFC) in Greenbelt, Maryland, USA. The data used in the present article is publicly available through NASA's {\it HEASARC} archive. We would like to thank Edo Berger, Bradley Cenko, Heino Falcke, Glennys Farrar,  Rob Fender,  Elmar K\"ording, Andrei Gruzinov, Jeroen Homan, Christian Knigge,  Julian Krolik, Sara Markoff, Andrea Merloni, Giulia Milgori, James Miller-Jones,  Gabriele Ponti, and especially Alan Marscher and Abdu Zoghbi for valuable discussions. We would like to thank the referee for detailed comments and suggestions that helped us improve the manuscript. Pasham is supported by NASA through an Einstein fellowship (PF6-170156) and van~Velzen is supported by NASA through a Hubble Fellowship (HST-HF2-51350). All the data presented here is public and can be found in Table~\ref{tab:xlong},  \citet{vanvelzen2016}, \citet{alexander2016}, and the NASA/{\it Swift} archive. \\*
Facilities: \facility{AMI}, \facility{VLA}, \facility{Swift}.

\bibliographystyle{apj} 
\bibliography{references.bib,general_desk.bib}

%\appendix
\setcounter{table}{0}
\renewcommand{\thetable}{A\arabic{table}}
\input{table3_rev2.tex}

\end{document}

%% file: table_jetfit_newsub.tex
\begin{deluxetable*}{c  c   r r r}

\tablewidth{0pt}
\tablecolumns{5}

\tablecaption{Best-fit parameters of an adiabatic jet model.}

\tablehead{Relevant equation 																								&Free parameters & $\phi=1/5$ 		& $\phi=1/10$ & $\phi=1/15$ d}
\startdata 
 \multirow{2}{*}{$B(z)=B_{0} (z/10^{15} {\rm cm})^{c_{B}}$} 		&	$B_{0}~({\rm G})$									&$9.6\,(2.3)$ 			&$7.6\,(2.0)$ 		&$8.0\,(2.0)$ 				\\
  																																	&	$c_{B}$  														&$-1.02\,(0.06)$		&$-1.02\,(0.05)$	&$-1.07\,(0.05)$			\\[3pt]				
 $Q_{\rm jet} \propto (z/z_{\rm head})^{c_{Q}}$ 								&	$c_{Q}$ 													 	&$1.20\,(0.4)$ 		 	&$1.20\,(0.4)$ 		&$1.40\,(0.2)$ 				\\[3pt]				
 																																	& $\log_{10}\gamma_{\rm max}$		&$1.9\,(0.4) $			&$2.0\,(0.3) $		&$2.0\,(0.2) $				\\[3pt]
 \multirow{2}{*}{$z_{\rm jet}(t) = v_{\rm jet}(t-t_{0})$}						&	$t_{0}~({\rm MJD}+56887)$				&$63\,(11)$ 				&$64\,(11)$ 			&$48\,(12)$ 					\\
 																										 							&	$v_{\rm jet}~(c)$									 	&$0.37\,(0.1)$ 		 	&$0.62\,(0.2)$ 		&$0.71\,(0.2)$ 				\\[3pt]

\cutinhead{Inferred parameters~~~~~~~~~~~~~~~}

%$\tau(16\,{\rm GHz})= \kappa z_{\rm laf}=1$										& $z_{\rm lag}/10^{16}~{\rm cm}$&$3.7\,(0.2)$ 		& $2.3\,(0.3)$ 			& $3.9\,(0.6)$ \\[3pt] % tau-weighted
$\tau(16\,{\rm GHz})= \kappa z_{\rm 16}=1$										& $z_{\rm 16}~(\times 10^{16}~{\rm cm})$
																																																				& $ 3.3_{-2.0}^{+2.1}$ & $3.4_{-1.6}^{+3.9}$ 	& $ 1.5_{-12}^{+2.8}$ \\[3pt]
{$v_{\rm lag}=( c\, \tau_{\rm lag}/ z_{\rm 16}+\cos i )^{-1}/\gamma $}	& $v_{\rm lag}~(c)$	
 																																																				& $0.6_{-0.2}^{+0.3}$	& $0.6_{-0.2}^{+0.3}$ 	& $0.3_{-0.2}^{+0.3}$\\[3pt]

\enddata
\tablecomments{The first six rows of the second column list the parameters of our jet model, with the relevant equations listed in the first column (see Sec.~\ref{sec:syncmodel} for details). Their best-fit values (including 1$\sigma$ uncertainties) are listed in the third to last columns, for three different values of the jet opening angle ($\phi$). We can use the best-fit parameters to infer the radius where the jet becomes optically thin to self-absorption at 16~GHz ($z_{\rm 16}$). Using the observed lag between the radio and the x-ray light curves ($\tau_{\rm lag}$) we can compute the mean communication velocity ($v_{\rm lag}$) between the black hole accretion disk and the 16~GHz self-absorption radius. We find that this estimate of the disk-jet communication velocity is consistent with the best-fit jet velocity ($v_{\rm jet}$). All parameters listed here are estimated after adopting a jet inclination of $i=60^{\circ}$.}\label{tab:jetmodel}
\end{deluxetable*}

%% file: table3_rev2.tex
\clearpage
\appendix
\LongTables
\begin{deluxetable}{ccccccc}
\centering
\tablecolumns{5}
\tablewidth{300pt}
\tablecaption{X-ray flux and count rates.}
\tablehead{ObsID$^{a}$ & {MJD$^{b}$} & {X-ray flux$^{c}$} & {X-ray count rate$^{d}$} & Exposure$^{e}$ & Pile-up radius$^{f}$ & {$\chi^2$/dof}$^{g}$}
\startdata
00033539001 & 56991.483 & 1.376$\pm$0.136 & 0.286$\pm$0.010 & 2952 & 5 & 1.34/21 \\
00033539002 & 56993.918 & 1.809$\pm$0.201 & 0.320$\pm$0.011 & 2812 & 6 & 1.11/18 \\
00033539003 & 56995.312 & 1.056$\pm$0.156 & 0.260$\pm$0.009 & 2997 & 7 & 1.10/10 \\
00033539004 & 56998.260 & 1.504$\pm$0.118 & 0.270$\pm$0.010 & 2952 & 4 & 0.77/28 \\
00033539005 & 57001.638 & 1.725$\pm$0.138 & 0.316$\pm$0.010 & 2892 & 4 & 1.43/25 \\
00033539006 & 57004.128 & 1.515$\pm$0.155 & 0.278$\pm$0.015 & 1179 & 7 & -/- \\
00033539007 & 57007.297 & 1.881$\pm$0.138 & 0.386$\pm$0.011 & 2944 & 4 & 0.76/29 \\
00033539008 & 57010.836 & 1.841$\pm$0.172 & 0.406$\pm$0.012 & 3044 & 6 & 1.20/21 \\
00033539009 & 57013.100 & 1.849$\pm$0.135 & 0.350$\pm$0.011 & 3122 & 4 & 1.50/28 \\
00033539010 & 57016.093 & 1.837$\pm$0.172 & 0.247$\pm$0.009 & 2829 & 5 & 0.89/23 \\
00033539011 & 57019.577 & 1.694$\pm$0.111 & 0.373$\pm$0.011 & 3169 & 3 & 1.60/28 \\
00033539012 & 57022.746 & 1.419$\pm$0.128 & 0.248$\pm$0.009 & 2764 & 4 & 0.73/23 \\
00033539014 & 57029.583 & 1.567$\pm$0.092 & 0.339$\pm$0.009 & 3773 & 3 & 1.38/35 \\
00033539015 & 57033.145 & 1.515$\pm$0.180 & 0.387$\pm$0.015 & 1681 & 5 & 1.09/14 \\
00033539016 & 57036.111 & 1.739$\pm$0.139 & 0.379$\pm$0.014 & 2075 & 3 & 1.13/26 \\
00033539017 & 57039.235 & 1.440$\pm$0.090 & 0.314$\pm$0.011 & 2797 & 2 & 0.77/33 \\
00033539018 & 57042.297 & 1.369$\pm$0.106 & 0.328$\pm$0.013 & 1943 & 2 & 1.61/29 \\
00033539019 & 57045.625 & 1.358$\pm$0.101 & 0.306$\pm$0.011 & 2325 & 2 & 1.09/28 \\
00033539020 & 57048.822 & 1.263$\pm$0.180 & 0.310$\pm$0.011 & 2489 & 7 & 1.32/11 \\
00033539021 & 57051.535 & 1.173$\pm$0.185 & 0.267$\pm$0.011 & 2342 & 7 & 1.78/11 \\
00033539022 & 57054.140 & 1.124$\pm$0.128 & 0.268$\pm$0.010 & 2909 & 5 & 1.07/16 \\
00033539023 & 57057.561 & 1.154$\pm$0.138 & 0.307$\pm$0.010 & 2909 & 6 & 0.43/14 \\
00033539024 & 57060.166 & 0.968$\pm$0.146 & 0.261$\pm$0.010 & 2565 & 7 & 1.46/10 \\
00033539025 & 57065.850 & 0.924$\pm$0.143 & 0.206$\pm$0.011 & 1831 & 5 & 0.55/9 \\
00033539026 & 57068.771 & 0.780$\pm$0.121 & 0.209$\pm$0.011 & 1581 & 7 & -/- \\
00033539027 & 57071.738 & 0.999$\pm$0.140 & 0.267$\pm$0.010 & 2482 & 6 & 0.98/11 \\
00033539028 & 57074.911 & 0.885$\pm$0.126 & 0.220$\pm$0.009 & 2847 & 6 & 0.39/10 \\
00033539029 & 57077.634 & 0.994$\pm$0.136 & 0.127$\pm$0.007 & 2470 & 5 & 1.08/12 \\
00033539030 & 57081.190 & 0.948$\pm$0.138 & 0.219$\pm$0.009 & 3012 & 6 & 0.88/10 \\
00033539032 & 57086.918 & 0.958$\pm$0.161 & 0.238$\pm$0.010 & 2215 & 6 & 1.73/9 \\
00033539033 & 57089.382 & 1.244$\pm$0.135 & 0.216$\pm$0.009 & 2857 & 5 & 0.70/18 \\
00033539034 & 57099.388 & 1.379$\pm$0.192 & 0.290$\pm$0.014 & 1471 & 6 & -/- \\
00033539035 & 57102.656 & 1.004$\pm$0.122 & 0.211$\pm$0.010 & 2118 & 4 & 0.97/14 \\
00033539036 & 57105.308 & 0.548$\pm$0.092 & 0.115$\pm$0.012 & 759 & 4 & -/- \\
00033539037 & 57108.768 & 0.918$\pm$0.151 & 0.223$\pm$0.010 & 2223 & 7 & -/- \\
00033539038 & 57111.934 & 0.926$\pm$0.141 & 0.225$\pm$0.010 & 2442 & 6 & 0.85/10 \\
00033539039 & 57114.090 & 1.000$\pm$0.166 & 0.243$\pm$0.012 & 1591 & 7 & -/- \\
00033539040 & 57117.683 & 0.721$\pm$0.127 & 0.221$\pm$0.010 & 2258 & 7 & -/- \\
00033539041 & 57120.318 & 0.701$\pm$0.115 & 0.215$\pm$0.010 & 2035 & 5 & 0.52/8 \\
00033539042 & 57123.540 & 0.652$\pm$0.120 & 0.174$\pm$0.009 & 2225 & 5 & -/- \\
00033539043 & 57126.244 & 0.682$\pm$0.116 & 0.182$\pm$0.009 & 2397 & 5 & 0.70/7 \\
00033539045 & 57129.401 & 0.799$\pm$0.084 & 0.196$\pm$0.011 & 1768 & 0 & 0.58/18 \\
00033539046 & 57132.561 & 0.578$\pm$0.075 & 0.143$\pm$0.008 & 2213 & 2 & 0.89/13 \\
00033539047 & 57136.561 & 0.585$\pm$0.077 & 0.162$\pm$0.011 & 1443 & 0 & 1.10/12 \\
00033539048 & 57139.348 & 0.837$\pm$0.100 & 0.153$\pm$0.008 & 2223 & 2 & 0.83/13 \\
00033539049 & 57147.598 & 0.712$\pm$0.082 & 0.142$\pm$0.008 & 2317 & 2 & 1.29/15 \\
00033539050 & 57150.257 & 0.817$\pm$0.066 & 0.183$\pm$0.008 & 2642 & 0 & 0.79/26 \\
00033539051 & 57153.487 & 0.741$\pm$0.068 & 0.177$\pm$0.008 & 2490 & 1 & 1.01/22 \\
00033539052 & 57156.349 & 0.666$\pm$0.080 & 0.159$\pm$0.010 & 1656 & 4 & -/- \\
00033539053 & 57173.107 & 0.656$\pm$0.068 & 0.156$\pm$0.009 & 1945 & 0 & 0.67/18 \\
00033539054 & 57176.134 & 0.630$\pm$0.070 & 0.177$\pm$0.010 & 1845 & 0 & 1.18/15 \\
00033539055 & 57179.029 & 0.498$\pm$0.080 & 0.140$\pm$0.014 & 691 & 2 & -/- \\
00033539056 & 57182.466 & 0.630$\pm$0.090 & 0.136$\pm$0.011 & 1134 & 0 & 0.63/10 \\
00033539057 & 57186.056 & 0.539$\pm$0.068 & 0.115$\pm$0.008 & 1666 & 0 & 1.29/13 \\
00033539059 & 57191.846 & 0.602$\pm$0.065 & 0.133$\pm$0.008 & 2160 & 0 & 0.60/18 \\
00033539060 & 57195.196 & 0.592$\pm$0.070 & 0.121$\pm$0.007 & 2298 & 2 & 1.07/15 \\
00033539061 & 57200.383 & 0.532$\pm$0.082 & 0.132$\pm$0.010 & 1301 & 1 & 0.96/9 \\
00033539062 & 57203.808 & 0.424$\pm$0.055 & 0.134$\pm$0.009 & 1836 & 0 & 0.55/11 \\
00033539063 & 57226.618 & 0.387$\pm$0.054 & 0.120$\pm$0.008 & 2008 & 0 & 0.74/11 \\
00033539064 & 57230.374 & 0.423$\pm$0.075 & 0.132$\pm$0.012 & 954 & 0 & -/- \\
00033539065 & 57236.471 & 0.404$\pm$0.059 & 0.124$\pm$0.007 & 2352 & 2 & 0.43/9 \\
00033539067 & 57242.121 & 0.452$\pm$0.053 & 0.106$\pm$0.007 & 2427 & 0 & 1.08/14 \\
00033539068 & 57246.908 & 0.342$\pm$0.047 & 0.075$\pm$0.006 & 2382 & 0 & 0.54/11 \\
00033539069 & 57340.746 & 0.397$\pm$0.051 & 0.063$\pm$0.004 & 3251 & 0 & 0.66/12 \\
00033539070 & 57351.806 & 0.247$\pm$0.054 & 0.050$\pm$0.005 & 2155 & 0 & -/- \\
00033539071 & 57354.699 & 0.227$\pm$0.039 & 0.046$\pm$0.004 & 2472 & 0 & 1.43/8 \\
00033539072 & 57357.365 & 0.322$\pm$0.048 & 0.066$\pm$0.005 & 2442 & 0 & 0.41/8 \\
00033539073 & 57360.259 & 0.362$\pm$0.055 & 0.090$\pm$0.007 & 1973 & 0 & 0.61/9 \\
00033539074 & 57363.952 & 0.366$\pm$0.048 & 0.096$\pm$0.006 & 2470 & 0 & 0.41/10 \\
00033539075 & 57366.941 & 0.393$\pm$0.054 & 0.073$\pm$0.005 & 2492 & 0 & 1.37/11 \\
00033539076 & 57369.828 & 0.370$\pm$0.064 & 0.069$\pm$0.005 & 2457 & 2 & -/- \\
00033539077 & 57372.213 & 0.331$\pm$0.060 & 0.062$\pm$0.006 & 1898 & 0 & -/- \\
00033539078 & 57375.538 & 0.285$\pm$0.059 & 0.061$\pm$0.006 & 1838 & 2 & -/- \\
00033539079 & 57378.412 & 0.358$\pm$0.059 & 0.077$\pm$0.006 & 1933 & 0 & 1.06/8 \\
00033539080 & 57383.201 & 0.133$\pm$0.042 & 0.029$\pm$0.007 & 531 & 0 & -/- \\
00033539082 & 57411.445 & 0.281$\pm$0.056 & 0.071$\pm$0.006 & 1766 & 2 & -/- \\
00033539084 & 57417.731 & 0.277$\pm$0.044 & 0.070$\pm$0.006 & 2125 & 0 & 0.94/8 \\
00033539085 & 57423.003 & 0.321$\pm$0.072 & 0.081$\pm$0.011 & 684 & 0 & -/- \\
00033539086 & 57426.465 & 0.157$\pm$0.048 & 0.040$\pm$0.010 & 419 & 0 & -/- \\
00033539087 & 57427.720 & 0.279$\pm$0.055 & 0.070$\pm$0.006 & 1988 & 0 & -/- \\
00033539088 & 57429.666 & 0.104$\pm$0.030 & 0.026$\pm$0.006 & 731 & 0 & -/- \\
00033539089 & 57432.788 & 0.194$\pm$0.040 & 0.049$\pm$0.005 & 1806 & 2 & -/- \\
00033539090 & 57435.706 & 0.199$\pm$0.041 & 0.050$\pm$0.005 & 1953 & 2 & -/- \\
00033539091 & 57519.777 & 0.182$\pm$0.038 & 0.046$\pm$0.005 & 2120 & 0 & -/- \\
00033539092 & 57522.502 & 0.139$\pm$0.029 & 0.035$\pm$0.004 & 2342 & 0 & -/- \\
00033539093 & 57526.691 & 0.142$\pm$0.030 & 0.036$\pm$0.004 & 2075 & 0 & -/- \\
00033539094 & 57542.648 & 0.128$\pm$0.029 & 0.032$\pm$0.004 & 1733 & 2 & -/- \\
00033539095 & 57545.454 & 0.127$\pm$0.040 & 0.032$\pm$0.008 & 474 & 0 & -/- \\
00033539096 & 57546.365 & 0.147$\pm$0.033 & 0.037$\pm$0.005 & 1441 & 2 & -/- \\
00033539097 & 57550.094 & 0.160$\pm$0.034 & 0.041$\pm$0.005 & 1818 & 2 & -/- \\
00033539098 & 57554.342 & 0.097$\pm$0.022 & 0.025$\pm$0.004 & 1998 & 2 & -/- \\
00033539099 & 57718.004 & 0.086$\pm$0.019 & 0.022$\pm$0.003 & 2410 & 0 & -/- \\
00033539100 & 57819.598 & 0.050$\pm$0.020 & 0.013$\pm$0.005 & 606 & 0 & -/- \\
00033539101 & 57820.399 & 0.055$\pm$0.014 & 0.014$\pm$0.003 & 1910 & 0 & -/- \\
00033539102 & 57821.936 & 0.096$\pm$0.023 & 0.024$\pm$0.004 & 1688 & 0 & -/- \\
00033539103 & 57826.844 & 0.029$\pm$0.010 & 0.007$\pm$0.002 & 1873 & 1 & -/- \\
00033539104 & 57828.563 & 0.071$\pm$0.021 & 0.018$\pm$0.004 & 1029 & 0 & -/- \\
00033539105 & 57833.492 & 0.089$\pm$0.025 & 0.022$\pm$0.005 & 954 & 0 & -/- \\
\enddata
\tablecomments{
$^{a}${\it Swift}-assigned observation IDs.$^{b}$Modified Julian Date. $^{c}$The X-ray fluxes were estimated from 0.3-1.0 keV bandpass and are in the units of 10$^{-11}$ erg s$^{-1}$ cm$^{-2}$. The uncertainties indicate both the lower and the upper 1$\sigma$ confidence levels. $^{d}$PSF-corrected X-ray count rates were also estimated from 0.3-1.0 keV bandpass and are in the units of counts per second. $^{e}$Exposure time in seconds. $^{f}$Inner exclusion radius in units of XRT pixels (1 pixel $\approx$ 2.36'') to mitigate photon pileup. This inner exclusion radius was determined by manually fitting the PSF in each exposure following the  methodology outlined in the {\it Swift}/XRT user guide (see sec. \ref{sec:xraydata}). $^{g}$The reduced $\chi^2$ along with the dof are indicated in the last column. For exposures marked by -/-, because the pile-up corrected counts were less than 100, the flux was estimated by scaling by the count rate of the nearest observation with spectral flux estimate (see sec. \ref{sec:xraydata}). \\
}\label{tab:xlong}
\end{deluxetable}
\clearpage

%% file: ms_final.bbl
\begin{thebibliography}{}
\expandafter\ifx\csname natexlab\endcsname\relax\def\natexlab#1{#1}\fi

\bibitem[{{Abramowicz} \& {Fragile}(2013)}]{Abramowicz13}
{Abramowicz}, M.~A., \& {Fragile}, P.~C. 2013, Living Reviews in Relativity,
  16, 1

\bibitem[{{Alexander} {et~al.}(2016){Alexander}, {Berger}, {Guillochon},
  {Zauderer}, \& {Williams}}]{alexander2016}
{Alexander}, K.~D., {Berger}, E., {Guillochon}, J., {Zauderer}, B.~A., \&
  {Williams}, P.~K.~G. 2016, \apjl, 819, L25

\bibitem[{{Alexander} {et~al.}(2017){Alexander}, {Wieringa}, {Berger},
  {Saxton}, \& {Komossa}}]{Alexander17}
{Alexander}, K.~D., {Wieringa}, M.~H., {Berger}, E., {Saxton}, R.~D., \&
  {Komossa}, S. 2017, \apj, 837, 153

\bibitem[{{Arcavi} {et~al.}(2014){Arcavi}, {Gal-Yam}, {Sullivan}, {Pan},
  {Cenko}, {Horesh}, {Ofek}, {De Cia}, {Yan}, {Yang}, {Howell}, {Tal},
  {Kulkarni}, {Tendulkar}, {Tang}, {Xu}, {Sternberg}, {Cohen}, {Bloom},
  {Nugent}, {Kasliwal}, {Perley}, {Quimby}, {Miller}, {Theissen}, \&
  {Laher}}]{Arcavi14}
{Arcavi}, I., {Gal-Yam}, A., {Sullivan}, M., {et~al.} 2014, \apj, 793, 38

\bibitem[{{Arnaud}(1996)}]{arnaud1996}
{Arnaud}, K.~A. 1996, in Astronomical Society of the Pacific Conference Series,
  Vol. 101, Astronomical Data Analysis Software and Systems V, ed. G.~H.
  {Jacoby} \& J.~{Barnes}, 17

\bibitem[{{Auchettl} {et~al.}(2017){Auchettl}, {Ramirez-Ruiz}, \&
  {Guillochon}}]{Auchettl17}
{Auchettl}, K., {Ramirez-Ruiz}, E., \& {Guillochon}, J. 2017, ArXiv e-prints,
  arXiv:1703.06141

\bibitem[{{Bade} {et~al.}(1996){Bade}, {Komossa}, \& {Dahlem}}]{Bade96}
{Bade}, N., {Komossa}, S., \& {Dahlem}, M. 1996, \aap, 309, L35

\bibitem[{{Bell} {et~al.}(2011){Bell}, {Tzioumis}, {Uttley}, {Fender},
  {Ar{\'e}valo}, {Breedt}, {McHardy}, {Calvelo}, {Jamil}, \&
  {K{\"o}rding}}]{bell2011}
{Bell}, M.~E., {Tzioumis}, T., {Uttley}, P., {et~al.} 2011, \mnras, 411, 402

\bibitem[{{Berger} {et~al.}(2012){Berger}, {Zauderer}, {Pooley}, {Soderberg},
  {Sari}, {Brunthaler}, \& {Bietenholz}}]{berger2012}
{Berger}, E., {Zauderer}, A., {Pooley}, G.~G., {et~al.} 2012, \apj, 748, 36

\bibitem[{{Blagorodnova} {et~al.}(2017){Blagorodnova}, {Gezari}, {Hung},
  {Kulkarni}, {Cenko}, {Pasham}, {Yan}, {Arcavi}, {Ben-Ami}, {Bue}, {Cantwell},
  {Cao}, {Castro-Tirado}, {Fender}, {Fremling}, {Gal-Yam}, {Ho}, {Horesh},
  {Hosseinzadeh}, {Kasliwal}, {Kong}, {Laher}, {Leloudas}, {Lunnan}, {Masci},
  {Mooley}, {Neill}, {Nugent}, {Powell}, {Valeev}, {Vreeswijk}, {Walters}, \&
  {Wozniak}}]{Blagorodnova17}
{Blagorodnova}, N., {Gezari}, S., {Hung}, T., {et~al.} 2017, ArXiv:1703.00965,
  arXiv:1703.00965

\bibitem[{{Blandford} \& {K{\"o}nigl}(1979)}]{blandfordkonigl1979}
{Blandford}, R.~D., \& {K{\"o}nigl}, A. 1979, \apj, 232, 34

\bibitem[{{Blandford} \& {Rees}(1974)}]{Blandford74}
{Blandford}, R.~D., \& {Rees}, M.~J. 1974, \mnras, 169, 395

\bibitem[{{Bloom} {et~al.}(2011){Bloom}, {Giannios}, {Metzger}, {Cenko},
  {Perley}, {Butler}, {Tanvir}, {Levan}, {O'Brien}, {Strubbe}, {De Colle},
  {Ramirez-Ruiz}, {Lee}, {Nayakshin}, {Quataert}, {King}, {Cucchiara},
  {Guillochon}, {Bower}, {Fruchter}, {Morgan}, \& {van der Horst}}]{Bloom2011}
{Bloom}, J.~S., {Giannios}, D., {Metzger}, B.~D., {et~al.} 2011, Science, 333,
  203

\bibitem[{{Bower} {et~al.}(2013){Bower}, {Metzger}, {Cenko}, {Silverman}, \&
  {Bloom}}]{Bower13}
{Bower}, G.~C., {Metzger}, B.~D., {Cenko}, S.~B., {Silverman}, J.~M., \&
  {Bloom}, J.~S. 2013, \apj, 763, 84

\bibitem[{{Bright} {et~al.}(2018){Bright}, {Fender}, {Motta}, {Mooley},
  {Perrott}, \& {van~Velzen}}]{Bright17}
{Bright}, J.~S., {Fender}, R.~P., {Motta}, S.~E., {et~al.} 2018, 
  \mnras, 475, 4011

\bibitem[{{Brown} {et~al.}(2015){Brown}, {Levan}, {Stanway}, {Tanvir}, {Cenko},
  {Berger}, {Chornock}, \& {Cucchiaria}}]{Brown15}
{Brown}, G.~C., {Levan}, A.~J., {Stanway}, E.~R., {et~al.} 2015, \mnras, 452,
  4297

\bibitem[{{Brown} {et~al.}(2017){Brown}, {Holoien}, {Auchettl}, {Stanek},
  {Kochanek}, {Shappee}, {Prieto}, \& {Grupe}}]{brownjs2017}
{Brown}, J.~S., {Holoien}, T.~W.-S., {Auchettl}, K., {et~al.} 2017, \mnras,
  466, 4904

\bibitem[{{Burbidge}(1956)}]{burbidge1956}
{Burbidge}, G.~R. 1956, \apj, 124, 416

\bibitem[{{Burrows} {et~al.}(2011){Burrows}, {Kennea}, {Ghisellini}, {Mangano},
  {Zhang}, {Page}, {Eracleous}, {Romano}, {Sakamoto}, {Falcone}, {Osborne},
  {Campana}, {Beardmore}, {Breeveld}, {Chester}, {Corbet}, {Covino},
  {Cummings}, {D'Avanzo}, {D'Elia}, {Esposito}, {Evans}, {Fugazza}, {Gelbord},
  {Hiroi}, {Holland}, {Huang}, {Im}, {Israel}, {Jeon}, {Jeon}, {Jun}, {Kawai},
  {Kim}, {Krimm}, {Marshall}, {P.~M{\'e}sz{\'a}ros}, {Negoro}, {Omodei},
  {Park}, {Perkins}, {Sugizaki}, {Sung}, {Tagliaferri}, {Troja}, {Ueda},
  {Urata}, {Usui}, {Antonelli}, {Barthelmy}, {Cusumano}, {Giommi}, {Melandri},
  {Perri}, {Racusin}, {Sbarufatti}, {Siegel}, \& {Gehrels}}]{burrows2011}
{Burrows}, D.~N., {Kennea}, J.~A., {Ghisellini}, G., {et~al.} 2011, \nat, 476,
  421

\bibitem[{{Cenko} {et~al.}(2012){Cenko}, {Krimm}, {Horesh}, {Rau}, {Frail},
  {Kennea}, {Levan}, {Holland}, {Butler}, {Quimby}, {Bloom}, {Filippenko},
  {Gal-Yam}, {Greiner}, {Kulkarni}, {Ofek}, {Olivares E.}, {Schady},
  {Silverman}, {Tanvir}, \& {Xu}}]{Cenko11}
{Cenko}, S.~B., {Krimm}, H.~A., {Horesh}, A., {et~al.} 2012, \apj, 753, 77

\bibitem[{{Chatterjee} {et~al.}(2011){Chatterjee}, {Marscher}, {Jorstad},
  {Markowitz}, {Rivers}, {Rothschild}, {McHardy}, {Aller}, {Aller},
  {L{\"a}hteenm{\"a}ki}, {Tornikoski}, {Harrison}, {Agudo}, {G{\'o}mez},
  {Taylor}, \& {Gurwell}}]{chatterjee2011}
{Chatterjee}, R., {Marscher}, A.~P., {Jorstad}, S.~G., {et~al.} 2011, \apj,
  734, 43

\bibitem[{{Chevalier}(1998)}]{chevalier1998}
{Chevalier}, R.~A. 1998, \apj, 499, 810

\bibitem[{{Coughlin} \& {Begelman}(2014)}]{Coughlin14}
{Coughlin}, E.~R., \& {Begelman}, M.~C. 2014, \apj, 781, 82

\bibitem[{{Croston} {et~al.}(2005){Croston}, {Hardcastle}, {Harris}, {Belsole},
  {Birkinshaw}, \& {Worrall}}]{croston2005}
{Croston}, J.~H., {Hardcastle}, M.~J., {Harris}, D.~E., {et~al.} 2005, \apj,
  626, 733

\bibitem[{{Crumley} {et~al.}(2017){Crumley}, {Ceccobello}, {Connors}, \&
  {Cavecchi}}]{crumley2017}
{Crumley}, P., {Ceccobello}, C., {Connors}, R.~M.~T., \& {Cavecchi}, Y. 2017,
  ArXiv e-prints, arXiv:1703.02842

\bibitem[{{Dai} {et~al.}(2015){Dai}, {McKinney}, \& {Miller}}]{Dai15}
{Dai}, L., {McKinney}, J.~C., \& {Miller}, M.~C. 2015, \apjl, 812, L39

\bibitem[{{Esquej} {et~al.}(2008){Esquej}, {Saxton}, {Komossa}, {Read},
  {Freyberg}, {Hasinger}, {Garc{\'{\i}}a-Hern{\'a}ndez}, {Lu}, {Rodriguez
  Zaur{\'{\i}}n}, {S{\'a}nchez-Portal}, \& {Zhou}}]{Esquej2008}
{Esquej}, P., {Saxton}, R.~D., {Komossa}, S., {et~al.} 2008, \aap, 489, 543

\bibitem[{{Falcke} \& {Biermann}(1995)}]{FB1995}
{Falcke}, H., \& {Biermann}, P.~L. 1995, \aap, 293, 665

\bibitem[{{Falcke} {et~al.}(2004){Falcke}, {K{\"o}rding}, \&
  {Markoff}}]{falcke2004}
{Falcke}, H., {K{\"o}rding}, E., \& {Markoff}, S. 2004, \aap, 414, 895

\bibitem[{{Falcke} {et~al.}(1995){Falcke}, {Malkan}, \&
  {Biermann}}]{falckeII1995}
{Falcke}, H., {Malkan}, M.~A., \& {Biermann}, P.~L. 1995, \aap, 298, 375

\bibitem[{{Falcke} {et~al.}(1999){Falcke}, {Bower}, {Lobanov}, {Krichbaum},
  {Patnaik}, {Aller}, {Aller}, {Ter{\"a}sranta}, {Wright}, \&
  {Sandell}}]{falcke1999}
{Falcke}, H., {Bower}, G.~C., {Lobanov}, A.~P., {et~al.} 1999, \apjl, 514, L17

\bibitem[{{French} {et~al.}(2016){French}, {Arcavi}, \& {Zabludoff}}]{French16}
{French}, K.~D., {Arcavi}, I., \& {Zabludoff}, A. 2016, \apjl, 818, L21

\bibitem[{{Gallo} {et~al.}(2003){Gallo}, {Fender}, \& {Pooley}}]{gallo2003}
{Gallo}, E., {Fender}, R.~P., \& {Pooley}, G.~G. 2003, \mnras, 344, 60

\bibitem[{{Gaskell} \& {Peterson}(1987)}]{gaskell1987}
{Gaskell}, C.~M., \& {Peterson}, B.~M. 1987, \apjs, 65, 1

\bibitem[{{Generozov} {et~al.}(2017){Generozov}, {Mimica}, {Metzger}, {Stone},
  {Giannios}, \& {Aloy}}]{Generozov17}
{Generozov}, A., {Mimica}, P., {Metzger}, B.~D., {et~al.} 2017, \mnras, 464,
  2481

\bibitem[{{Gezari} {et~al.}(2009){Gezari}, {Heckman}, {Cenko}, {Eracleous},
  {Forster}, {Gon{\c c}alves}, {Martin}, {Morrissey}, {Neff}, {Seibert},
  {Schiminovich}, \& {Wyder}}]{Gezari09}
{Gezari}, S., {Heckman}, T., {Cenko}, S.~B., {et~al.} 2009, \apj, 698, 1367

\bibitem[{{Giannios} \& {Metzger}(2011)}]{Giannios11}
{Giannios}, D., \& {Metzger}, B.~D. 2011, \mnras, 416, 2102

\bibitem[{{Harwood} {et~al.}(2016){Harwood}, {Croston}, {Intema}, {Stewart},
  {Ineson}, {Hardcastle}, {Godfrey}, {Best}, {Brienza}, {Heesen}, {Mahony},
  {Morganti}, {Murgia}, {Orr{\'u}}, {R{\"o}ttgering}, {Shulevski}, \&
  {Wise}}]{hardwood2016}
{Harwood}, J.~J., {Croston}, J.~H., {Intema}, H.~T., {et~al.} 2016, \mnras,
  458, 4443

\bibitem[{{Heinz} \& {Sunyaev}(2003)}]{heinz_sunyaev03}
{Heinz}, S., \& {Sunyaev}, R.~A. 2003, \mnras, 343, L59

\bibitem[{{Herrnstein} {et~al.}(1997){Herrnstein}, {Moran}, {Greenhill},
  {Diamond}, {Miyoshi}, {Nakai}, \& {Inoue}}]{herrnstein1997}
{Herrnstein}, J.~R., {Moran}, J.~M., {Greenhill}, L.~J., {et~al.} 1997, \apjl,
  475, L17


\bibitem[{{Holoien} {et~al.}(2016){Holoien}, {Kochanek},
  {Prieto}, {Stanek}, {Dong}, {Shappee}, {Grupe}, {Brown}, {Basu}, {Beacom},
  {Bersier}, {Brimacombe}, {Danilet}, {Falco}, {Guo}, {Jose}, {Herczeg},
  {Long}, {Pojmanski}, {Simonian}, {Szczygie{\l}}, {Thompson}, {Thorstensen},
  {Wagner}, \& {Wo{\'z}niak}}]{Holoien16}
---. 2016{\natexlab{b}}, \mnras, 455, 2918

\bibitem[{{Irwin} {et~al.}(2015){Irwin}, {Henriksen}, {Krause}, {Wang},
  {Wiegert}, {Murphy}, {Heald}, \& {Perlman}}]{Irwin15}
{Irwin}, J.~A., {Henriksen}, R.~N., {Krause}, M., {et~al.} 2015, \apj, 809, 172

\bibitem[{{Kara} {et~al.}(2016){Kara}, {Miller}, {Reynolds}, \& {Dai}}]{Kara16}
{Kara}, E., {Miller}, J.~M., {Reynolds}, C., \& {Dai}, L. 2016, \nat, 535, 388

\bibitem[{{Kataoka} \& {Stawarz}(2005)}]{Kataoka2005}
{Kataoka}, J., \& {Stawarz}, {\L}. 2005, \apj, 622, 797

\bibitem[{{Ker} {et~al.}(2012){Ker}, {Best}, {Rigby}, {R{\"o}ttgering}, \&
  {Gendre}}]{Ker12}
{Ker}, L.~M., {Best}, P.~N., {Rigby}, E.~E., {R{\"o}ttgering}, H.~J.~A., \&
  {Gendre}, M.~A. 2012, \mnras, 420, 2644

\bibitem[{{Komossa}(2002)}]{komossa2002}
{Komossa}, S. 2002, in Reviews in Modern Astronomy, Vol.~15, Reviews in Modern
  Astronomy, ed. R.~E. {Schielicke}, 27

\bibitem[{{Krolik} {et~al.}(2016){Krolik}, {Piran}, {Svirski}, \&
  {Cheng}}]{krolik2016}
{Krolik}, J., {Piran}, T., {Svirski}, G., \& {Cheng}, R.~M. 2016, \apj, 827,
  127

\bibitem[{{Levan} {et~al.}(2011){Levan}, {Tanvir}, {Cenko}, {Perley},
  {Wiersema}, {Bloom}, {Fruchter}, {Postigo}, {O'Brien}, {Butler}, {van der
  Horst}, {Leloudas}, {Morgan}, {Misra}, {Bower}, {Farihi}, {Tunnicliffe},
  {Modjaz}, {Silverman}, {Hjorth}, {Th{\"o}ne}, {Cucchiara}, {Cer{\'o}n},
  {Castro-Tirado}, {Arnold}, {Bremer}, {Brodie}, {Carroll}, {Cooper}, {Curran},
  {Cutri}, {Ehle}, {Forbes}, {Fynbo}, {Gorosabel}, {Graham}, {Hoffman},
  {Guziy}, {Jakobsson}, {Kamble}, {Kerr}, {Kasliwal}, {Kouveliotou},
  {Kocevski}, {Law}, {Nugent}, {Ofek}, {Poznanski}, {Quimby}, {Rol},
  {Romanowsky}, {S{\'a}nchez-Ram{\'{\i}}rez}, {Schulze}, {Singh}, {van
  Spaandonk}, {Starling}, {Strom}, {Tello}, {Vaduvescu}, {Wheatley}, {Wijers},
  {Winters}, \& {Xu}}]{Levan11}
{Levan}, A.~J., {Tanvir}, N.~R., {Cenko}, S.~B., {et~al.} 2011, Science, 333,
  199

\bibitem[{{Levan} {et~al.}(2016){Levan}, {Tanvir}, {Brown}, {Metzger}, {Page},
  {Cenko}, {O'Brien}, {Lyman}, {Wiersema}, {Stanway}, {Fruchter}, {Perley}, \&
  {Bloom}}]{Levan16}
{Levan}, A.~J., {Tanvir}, N.~R., {Brown}, G.~C., {et~al.} 2016, \apj, 819, 51

\bibitem[{{Lodato} {et~al.}(2009){Lodato}, {King}, \& {Pringle}}]{Lodato09}
{Lodato}, G., {King}, A.~R., \& {Pringle}, J.~E. 2009, \mnras, 392, 332

\bibitem[{{Lodato} \& {Rossi}(2011)}]{Lodato11}
{Lodato}, G., \& {Rossi}, E.~M. 2011, \mnras, 410, 359

\bibitem[{{Markoff} {et~al.}(2005){Markoff}, {Nowak}, \& {Wilms}}]{markoff2005}
{Markoff}, S., {Nowak}, M.~A., \& {Wilms}, J. 2005, \apj, 635, 1203

\bibitem[{{Marscher}(1980)}]{Marscher1980}
{Marscher}, A.~P. 1980, \apj, 235, 386

\bibitem[{{Marscher} \& {Gear}(1985)}]{marscher1985}
{Marscher}, A.~P., \& {Gear}, W.~K. 1985, \apj, 298, 114

\bibitem[{{Marscher} {et~al.}(2002){Marscher}, {Jorstad}, {G{\'o}mez}, {Aller},
  {Ter{\"a}sranta}, {Lister}, \& {Stirling}}]{marscher2002}
{Marscher}, A.~P., {Jorstad}, S.~G., {G{\'o}mez}, J.-L., {et~al.} 2002, \nat,
  417, 625

\bibitem[{{Merloni} {et~al.}(2003){Merloni}, {Heinz}, \& {di
  Matteo}}]{merloni2003}
{Merloni}, A., {Heinz}, S., \& {di Matteo}, T. 2003, \mnras, 345, 1057

\bibitem[{{Metzger} {et~al.}(2012){Metzger}, {Giannios}, \&
  {Mimica}}]{Metzger12}
{Metzger}, B.~D., {Giannios}, D., \& {Mimica}, P. 2012, \mnras, 420, 3528

\bibitem[{{M{\"i}ller} \& {G{\"u}ltekin}(2011)}]{millersw1644radio2011}
{M{\"i}ller}, J.~M., \& {G{\"u}ltekin}, K. 2011, \apjl, 738, L13

\bibitem[{{Miller} {et~al.}(2015){Miller}, {Kaastra}, {Miller}, {Reynolds},
  {Brown}, {Cenko}, {Drake}, {Gezari}, {Guillochon}, {Gultekin}, {Irwin},
  {Levan}, {Maitra}, {Maksym}, {Mushotzky}, {O'Brien}, {Paerels}, {de Plaa},
  {Ramirez-Ruiz}, {Strohmayer}, \& {Tanvir}}]{miller2015}
{Miller}, J.~M., {Kaastra}, J.~S., {Miller}, M.~C., {et~al.} 2015, \nat, 526,
  542

\bibitem[{{Mimica} {et~al.}(2015){Mimica}, {Giannios}, {Metzger}, \&
  {Aloy}}]{Mimica15}
{Mimica}, P., {Giannios}, D., {Metzger}, B.~D., \& {Aloy}, M.~A. 2015, \mnras,
  450

\bibitem[{{Nakar} \& {Piran}(2011)}]{Nakar11}
{Nakar}, E., \& {Piran}, T. 2011, \nat, 478, 82

\bibitem[{{Pacholczyk}(1970)}]{Pacholczyk70}
{Pacholczyk}, A.~G. 1970, {Radio astrophysics. Nonthermal processes in galactic
  and extragalactic sources} ({Freeman and Co., San Francisco})

\bibitem[{{Pasham} {et~al.}(2017){Pasham}, {Cenko}, {Sadowski}, {Guillochon},
  {Stone}, {van Velzen}, \& {Cannizzo}}]{Pasham17}
{Pasham}, D.~R., {Cenko}, S.~B., {Sadowski}, A., {et~al.} 2017, \apjl, 837, L30

\bibitem[{{Pasham} {et~al.}(2015){Pasham}, {Cenko}, {Levan}, {Bower}, {Horesh},
  {Brown}, {Dolan}, {Wiersema}, {Filippenko}, {Fruchter}, {Greiner}, {O'Brien},
  {Page}, {Rau}, \& {Tanvir}}]{Pasham15}
{Pasham}, D.~R., {Cenko}, S.~B., {Levan}, A.~J., {et~al.} 2015, \apj, 805, 68

\bibitem[{{Perlman} {et~al.}(2017){Perlman}, {Meyer}, {Wang}, {Yuan},
  {Henriksen}, {Irwin}, {Krause}, {Wiegert}, {Murphy}, {Heald}, \&
  {Dettmar}}]{Perlman17}
{Perlman}, E.~S., {Meyer}, E.~T., {Wang}, Q.~D., {et~al.} 2017, \apj, 842, 126

\bibitem[{{Peterson} {et~al.}(2004){Peterson}, {Ferrarese}, {Gilbert}, {Kaspi},
  {Malkan}, {Maoz}, {Merritt}, {Netzer}, {Onken}, {Pogge}, {Vestergaard}, \&
  {Wandel}}]{peterson2004}
{Peterson}, B.~M., {Ferrarese}, L., {Gilbert}, K.~M., {et~al.} 2004, \apj, 613,
  682

\bibitem[{{Phinney}(1989)}]{phinney1989}
{Phinney}, E.~S. 1989, in IAU Symposium, Vol. 136, The Center of the Galaxy,
  ed. M.~{Morris}, 543

\bibitem[{{Piran}(2004)}]{Piran04}
{Piran}, T. 2004, Reviews of Modern Physics, 76, 1143

\bibitem[{{Prieto} {et~al.}(2016){Prieto}, {Kr{\"u}hler}, {Anderson},
  {Galbany}, {Kochanek}, {Aquino}, {Brown}, {Dong}, {F{\"o}rster}, {Holoien},
  {Kuncarayakti}, {Maureira}, {Rosales-Ortega}, {S{\'a}nchez}, {Shappee}, \&
  {Stanek}}]{Prieto16}
{Prieto}, J.~L., {Kr{\"u}hler}, T., {Anderson}, J.~P., {et~al.} 2016, \apjl,
  830, L32

\bibitem[{{Rawlings} \& {Saunders}(1991)}]{Rawlings91}
{Rawlings}, S., \& {Saunders}, R. 1991, \nat, 349, 138

\bibitem[{{Readhead} {et~al.}(1978){Readhead}, {Cohen}, {Pearson}, \&
  {Wilkinson}}]{readhead1978}
{Readhead}, A.~C.~S., {Cohen}, M.~H., {Pearson}, T.~J., \& {Wilkinson}, P.~N.
  1978, \nat, 276, 768

\bibitem[{{Rees}(1988)}]{rees1988}
{Rees}, M.~J. 1988, \nat, 333, 523

\bibitem[{{Remillard} \& {McClintock}(2006)}]{RemillardMcClintock06}
{Remillard}, R.~A., \& {McClintock}, J.~E. 2006, \araa, 44, 49

\bibitem[{{Romero-Ca{\~n}izales} {et~al.}(2011){Romero-Ca{\~n}izales},
  {Mattila}, {Alberdi}, {P{\'e}rez-Torres}, {Kankare}, \&
  {Ryder}}]{Romero_variability}
{Romero-Ca{\~n}izales}, C., {Mattila}, S., {Alberdi}, A., {et~al.} 2011,
  \mnras, 415, 2688

\bibitem[{{Romero-Ca{\~n}izales} {et~al.}(2016){Romero-Ca{\~n}izales},
  {Prieto}, {Chen}, {Kochanek}, {Dong}, {Holoien}, {Stanek}, \&
  {Liu}}]{14lievn2016}
{Romero-Ca{\~n}izales}, C., {Prieto}, J.~L., {Chen}, X., {et~al.} 2016, \apjl,
  832, L10

\bibitem[{{Saxton} {et~al.}(2017){Saxton}, {Read}, {Komossa}, {Lira},
  {Alexander}, \& {Wieringa}}]{Saxton17}
{Saxton}, R.~D., {Read}, A.~M., {Komossa}, S., {et~al.} 2017, \aap, 598, A29

\bibitem[{{S{\c a}dowski} {et~al.}(2011){S{\c a}dowski}, {Bursa}, {Abramowicz},
  {Klu{\'z}niak}, {Lasota}, {Moderski}, \& {Safarzadeh}}]{Sadowski11}
{S{\c a}dowski}, A., {Bursa}, M., {Abramowicz}, M., {et~al.} 2011, \aap, 532,
  A41

\bibitem[{{S{\c a}dowski} \& {Narayan}(2016)}]{olekdiskwind}
{S{\c a}dowski}, A., \& {Narayan}, R. 2016, \mnras, 456, 3929

\bibitem[{{Shappee} {et~al.}(2014){Shappee}, {Prieto}, {Grupe}, {Kochanek},
  {Stanek}, {De Rosa}, {Mathur}, {Zu}, {Peterson}, {Pogge}, {Komossa}, {Im},
  {Jencson}, {Holoien}, {Basu}, {Beacom}, {Szczygie{\l}}, {Brimacombe},
  {Adams}, {Campillay}, {Choi}, {Contreras}, {Dietrich}, {Dubberley},
  {Elphick}, {Foale}, {Giustini}, {Gonzalez}, {Hawkins}, {Howell}, {Hsiao},
  {Koss}, {Leighly}, {Morrell}, {Mudd}, {Mullins}, {Nugent}, {Parrent},
  {Phillips}, {Pojmanski}, {Rosing}, {Ross}, {Sand}, {Terndrup}, {Valenti},
  {Walker}, \& {Yoon}}]{shappee2014}
{Shappee}, B.~J., {Prieto}, J.~L., {Grupe}, D., {et~al.} 2014, \apj, 788, 48

\bibitem[{{Tchekhovskoy} {et~al.}(2014){Tchekhovskoy}, {Metzger}, {Giannios},
  \& {Kelley}}]{Tchekhovskoy13}
{Tchekhovskoy}, A., {Metzger}, B.~D., {Giannios}, D., \& {Kelley}, L.~Z. 2014,
  \mnras, 437, 2744

\bibitem[{{Tremaine} {et~al.}(2002){Tremaine}, {Gebhardt}, {Bender}, {Bower},
  {Dressler}, {Faber}, {Filippenko}, {Green}, {Grillmair}, {Ho}, {Kormendy},
  {Lauer}, {Magorrian}, {Pinkney}, \& {Richstone}}]{tremaine2002}
{Tremaine}, S., {Gebhardt}, K., {Bender}, R., {et~al.} 2002, \apj, 574, 740

\bibitem[{{van der Laan}(1966)}]{vanderlaan1966}
{van der Laan}, H. 1966, \nat, 211, 1131

\bibitem[{{van Velzen} {et~al.}(2015){van Velzen}, {Falcke}, \&
  {K{\"o}rding}}]{vanVelzen15}
{van Velzen}, S., {Falcke}, H., \& {K{\"o}rding}, E. 2015, \mnras, 446, 2985

\bibitem[{{van Velzen} {et~al.}(2013){van Velzen}, {Frail}, {K{\"o}rding}, \&
  {Falcke}}]{vanVelzen12b}
{van Velzen}, S., {Frail}, D.~A., {K{\"o}rding}, E., \& {Falcke}, H. 2013,
  \aap, 552, A5

\bibitem[{van Velzen {et~al.}(2011)van Velzen, K\"ording, \&
  Falcke}]{vanVelzen11}
van Velzen, S., K\"ording, E., \& Falcke, H. 2011, \mnras, 417, L51

\bibitem[{{van Velzen} {et~al.}(2016{\natexlab{a}}){van Velzen}, {Mendez},
  {Krolik}, \& {Gorjian}}]{vanVelzen16b}
{van Velzen}, S., {Mendez}, A.~J., {Krolik}, J.~H., \& {Gorjian}, V.
  2016{\natexlab{a}}, \apj, 829, 19

\bibitem[{{van Velzen} {et~al.}(2011){van Velzen}, {Farrar}, {Gezari},
  {Morrell}, {Zaritsky}, {{\"O}stman}, {Smith}, {Gelfand}, \&
  {Drake}}]{vanVelzen10}
{van Velzen}, S., {Farrar}, G.~R., {Gezari}, S., {et~al.} 2011, \apj, 741, 73

\bibitem[{{van Velzen} {et~al.}(2016{\natexlab{b}}){van Velzen}, {Anderson},
  {Stone}, {Fraser}, {Wevers}, {Metzger}, {Jonker}, {van der Horst}, {Staley},
  {Mendez}, {Miller-Jones}, {Hodgkin}, {Campbell}, \& {Fender}}]{vanvelzen2016}
{van Velzen}, S., {Anderson}, G.~E., {Stone}, N.~C., {et~al.}
  2016{\natexlab{b}}, Science, 351, 62

\bibitem[{{Wevers} {et~al.}(2017){Wevers}, {van Velzen}, {Jonker}, {Stone},
  {Hung}, {Onori}, {Gezari}, \& {Blagorodnova}}]{Wevers17}
{Wevers}, T., {van Velzen}, S., {Jonker}, P.~G., {et~al.} 2017, \mnras, 471,
  1694

\bibitem[{{White} \& {Peterson}(1994)}]{white1994}
{White}, R.~J., \& {Peterson}, B.~M. 1994, \pasp, 106, 879

\bibitem[{{Yuan} \& {Cui}(2005)}]{yuancui2005}
{Yuan}, F., \& {Cui}, W. 2005, \apj, 629, 408

\bibitem[{{Zauderer} {et~al.}(2011){Zauderer}, {Berger}, {Soderberg}, {Loeb},
  {Narayan}, {Frail}, {Petitpas}, {Brunthaler}, {Chornock}, {Carpenter},
  {Pooley}, {Mooley}, {Kulkarni}, {Margutti}, {Fox}, {Nakar}, {Patel},
  {Volgenau}, {Culverhouse}, {Bietenholz}, {Rupen}, {Max-Moerbeck}, {Readhead},
  {Richards}, {Shepherd}, {Storm}, \& {Hull}}]{zauderer2011}
{Zauderer}, B.~A., {Berger}, E., {Soderberg}, A.~M., {et~al.} 2011, \nat, 476,
  425

\end{thebibliography}
